\documentclass[usenatbib]{mn2e}

\usepackage{longtable}
\usepackage{graphicx}
\usepackage{epsfig}
\newcommand{\ha}{H$\alpha$} 
\newcommand{\hi}{H{\sc i}} 
\newcommand{\hii}{H{\sc ii}} 
\newcommand{\paper}{Paper {\sc i}}
\newcommand{\farc}{\ensuremath{^{\prime}\!}}
\def\vhel{\ifmmode{V_{{\rm HEL}}}\else{$V_{{\rm HEL}}$}\fi}
\def\vsys{\ifmmode{V_{\rm sys}}\else{$V_{\rm sys}$}\fi}
\def\kms{\ifmmode{\,{\rm km\,s}^{-1}}\else{~km~s$^{-1}$}\fi}
\def\vlsr{\ifmmode{v_{\rm lsr}}\else{$v_{\rm lsr}$}\fi}

\def\ltsim{\ifmmode\stackrel{<}{_{\sim}}\else$\stackrel{<}{_{\sim}}$\fi}
\def\gtsim{\ifmmode\stackrel{>}{_{\sim}}\else$\stackrel{>}{_{\sim}}$\fi}

\def\reff@jnl#1{{\rm#1\/}}

\def\aj{\reff@jnl{AJ}}                  % Astronomical Journal
\def\araa{\reff@jnl{ARA\&A}}            % Annual Review of Astron and Astrophys
\def\apj{\reff@jnl{ApJ}}                % Astrophysical Journal
\def\apjl{\reff@jnl{ApJ}}               % Astrophysical Journal, Letters
\def\apjs{\reff@jnl{ApJS}}              % Astrophysical Journal, Supplement
\def\ao{\reff@jnl{Appl.Optics}}         % Applied Optics
\def\apss{\reff@jnl{Ap\&SS}}            % Astrophysics and Space Science
\def\aap{\reff@jnl{A\&A}}               % Astronomy and Astrophysics
\def\aapr{\reff@jnl{A\&A~Rev.}}         % Astronomy and Astrophysics Reviews
\def\aaps{\reff@jnl{A\&AS}}             % Astronomy and Astrophysics, Supplement
\def\azh{\reff@jnl{AZh}}                        % Astronomicheskii Zhurnal
\def\baas{\reff@jnl{BAAS}}              % Bulletin of the AAS
\def\jrasc{\reff@jnl{JRASC}}            % Journal of the RAS of Canada
\def\memras{\reff@jnl{MmRAS}}           % Memoirs of the RAS
\def\mnras{\reff@jnl{MNRAS}}            % Monthly Notices of the RAS
\def\pra{\reff@jnl{Phys.Rev.A}}         % Physical Review A: General Physics
\def\prb{\reff@jnl{Phys.Rev.B}}         % Physical Review B: Solid State
\def\prc{\reff@jnl{Phys.Rev.C}}         % Physical Review C
\def\prd{\reff@jnl{Phys.Rev.D}}         % Physical Review D
\def\prl{\reff@jnl{Phys.Rev.Lett}}      % Physical Review Letters
\def\pasp{\reff@jnl{PASP}}              % Publications of the ASP
\def\pasj{\reff@jnl{PASJ}}              % Publications of the ASJ
\def\qjras{\reff@jnl{QJRAS}}            % Quarterly Journal of the RAS
\def\skytel{\reff@jnl{S\&T}}            % Sky and Telescope
\def\solphys{\reff@jnl{Solar~Phys.}}    % Solar Physics
\def\sovast{\reff@jnl{Soviet~Ast.}}     % Soviet Astronomy
 \def\ssr{\reff@jnl{Space~Sci.Rev.}}     % Space Science Reviews
\def\zap{\reff@jnl{ZAp}}                        % Zeitschrift fuer Astrophysik
\def\nat{\reff@jnl{Nature}}             % Nature 

\def\LaTeX{L\kern-.36em\raise.3ex\hbox{a}\kern-.15em
    T\kern-.1667em\lower.7ex\hbox{E}\kern-.125emX}

\begin{document}

\title[The free-free emission on the Galactic plane]{A derivation of the free-free emission on the Galactic plane between $\ell=20$\degr~and $44$\degr}
\author[M.I.R. Alves et al.]{Marta I. R. Alves,$\!^{1,2}$\thanks{E-mail:marta.alves@ias.u-psud.fr} Rodney D. Davies,$\!^{1}$ Clive Dickinson,$\!^{1}$ Mark Calabretta,$\!^{3}$\newauthor Richard Davis,$\!^{1}$ Lister Staveley-Smith$^{4}$ \\
$^1$Jodrell Bank Centre for Astrophysics, Alan Turing Building, School of Physics and Astronomy, \\
The University of Manchester, Oxford Road, Manchester, M13 9PL, UK \\
$^2$Institut d'Astrophysique Spatiale, CNRS (8617) Universit\'{e} Paris-Sud 11, B\^{a}timent 121, Orsay, France\\
$^3$Australia Telescope National Facility, PO Box 76, Epping, NSW 1710, Australia\\
$^4$International Centre for Radio Astronomy Research, M468, University of Western Australia,\\
35 Stirling Hwy, Crawley, WA 6009, Australia\\
}

\date{Received **insert**; Accepted **insert**}
       
\pagerange{\pageref{firstpage}--\pageref{lastpage}} 
\pubyear{}

\maketitle
\label{firstpage}

%%%%%%%%%%%%%%%%%%%%%%%%%%%%%%%%%%%%%%%%%%%%%%%%%%%%%%%%%%%%%%%%%%%%%%%%%%

\begin{abstract}

We present the derivation of the free-free emission on the Galactic plane between $\ell=20$\degr~and $44$\degr~and $|b| \leq 4$\degr, using Radio Recombination Line (RRL) data from the \hi~Parkes All-Sky Survey (HIPASS). Following an upgrade on the RRL data reduction technique, which improves significantly the quality of the final RRL spectra, we have extended the analysis to three times the area covered in \citet{Alves:2010}. The final RRL map has an angular resolution of 14.8\,arcmin and a velocity resolution of 20\kms. 

A map of the electron temperature ($T_{\rm e}$) of the ionised gas is derived for the area under study using the line and continuum data from the present survey. The mean $T_{\rm e}$ on the Galactic plane is 6000\,K. The first direct measure of the free-free emission is obtained based on the derived $T_{\rm e}$ map. Subtraction of this thermal component from the total continuum leaves the first direct measure of the synchrotron emission at 1.4\,GHz. A narrow component of width 2\degr~is identified in the latitude distribution of the synchrotron emission. We present a list of \hii~regions and SNRs extracted from the present free-free and synchrotron maps, where we confirm the synchrotron nature of three objects: G41.12-0.21, G41.15+0.39 and G35.59-0.44. We also identify a bright (42~Jy) new double radio galaxy, J1841-0152, previously unrecognised owing to the high optical extinction in the region. 

The latitude distribution for the RRL-derived free-free emission shows that the WMAP Maximum Entropy Method (MEM) is too high by $\sim 50$ per cent, in agreement with other recent results. The extension of this study to the inner Galaxy region $\ell=-50^{\circ}$ to 50\degr~will allow a better overall comparison of the RRL result with WMAP.

\end{abstract}

\begin{keywords}
radiation mechanisms: general -- methods: data analysis -- dust, extinction -- \hii~regions -- ISM: lines and bands -- Galaxy structure -- radio lines: ISM

\end{keywords}
\setcounter{figure}{0}

%%%%%%%%%%%%%%%%%%%%%%%%%%%%%%%%%%%%%%%%%%%%%%%%%%%%%%%%%%%%%%%%%%%%%%%%%%
%%%%%%%%%%%%%%%%%%%%%%%%%%%%%%%%%%%%%%%%%%%%%%%%%%%%%%%%%%%%%%%%%%%%%%%%%%
\section{INTRODUCTION}
\label{sec:introduction}

Radio Recombination Lines (RRLs) are a powerful tool for the diagnostics of the interstellar medium, tracing the ionised component, its electron temperature, velocity and radial distributions. They can be used to determine the free-free Emission Measure on the Galactic plane where heavy obscuration makes optical measurements impossible. In \citet{Alves:2010}, hereafter referred to as \paper, we have presented the unique method of determining the free-free emission on the Galactic plane using RRLs. In the present work we include a determination of the electron temperature of the ionised gas, following the separation of the continuum free-free and synchrotron emission.

Once the free-free emission is determined, the synchrotron component is available by comparing with the total continuum emission at the same frequency, from the same survey. This approach is more rigorous than the normal method of using the different spectral indices of the two components to make the separation. It allows a first clear image of the Galactic latitude and longitude distributions of the two components.

\paper~showed the feasibility of the RRL approach. The aim of the present paper is to extend the work to a larger longitude range in order to sample a greater volume of the Galaxy. The original paper covered the Sagittarius arm in the first quadrant; here we include more of the Sagittarius arm and a Section of the Scutum arm. Separating the emission from the two spiral arms allows the investigation of the z-thickness of the ionised layer. Early determinations of the latitude distribution of the diffuse RRL emission on the Galactic plane gave a thin layer of scale height $40-100$~pc \citet{Gordon:1972,Hart:1976,Mezger:1978}. These values are in agreement with the z-distribution of the \hii~regions within the solar circle, with a FWHM (Full Width at Half Maximum) of $\sim 90-120$~pc \citep{Paladini:2004} and somewhat greater than that of the OB stars with a FWHM of 32~pc \citep{Bronfman:2000}.

The analysis of \paper~was based on the deep ZOA Galactic plane survey which made a first attempt at removing the spectral ripples in RRL work.
We improve the signal-to-noise of the spectra and the spatial ripple by including the HIPASS data which was acquired by scanning in declination, as opposed to Galactic longitude scanning in the ZOA survey. We describe the update of the RRL reduction pipeline used in \paper, which includes a new bandpass algorithm exclusively designed for the analysis of RRL spectral line data in HIPASS. It extends the analysis from the previous $8^{\circ}(\ell) \times 8^{\circ}(b)$ region centred at $(\ell,b)=(40^{\circ},0^{\circ})$ to a $24^{\circ}(\ell) \times 8^{\circ}(b)$ area centred at $(\ell,b)=(32^{\circ},0^{\circ})$.

The paper is organised as follows: Section \ref{sec:rrlsurv} gives the HIPASS/ZOA survey parameters. In Section \ref{sec:rrldata} we describe the RRL data reduction and the properties of the final data cube. The derivation of the free-free map is presented in Section \ref{sec:fftemp}, along with a catalogue of \hii~regions extracted from the same map. In Section \ref{sec:synctemp} we present the synchrotron map derived from this survey with its catalogue of SNRs. Section \ref{sec:mem} compares the RRL estimates of the free-free with \textit{WMAP} data. Finally, Section \ref{sec:3dgal} gives the velocity and radial distribution of the ionised gas in the Galaxy.

%%%%%%%%%%%%%%%%%%%%%%%%%%%%%%%%%%%%%%%%%%%%%%%%%%%%%%%%%%%%%%%%%%%%%%
%%%%%%%%%%%%%%%%%%%%%%%%%%%%%%%%%%%%%%%%%%%%%%%%%%%%%%%%%%%%%%%%%%%%%%
\section{THE RRL SURVEY}
\label{sec:rrlsurv}

The sensitive survey described in this work is a product of the \hi~Parkes All-Sky Survey (HIPASS, \citealt{Staveley-Smith:1996}) and associated Zone of Avoidance (ZOA) survey \citep{Staveley-Smith:1998}. The surveys searched for \hi-emitting galaxies in the redshift velocities of $-1280$ and $12700$\,\kms~using the Parkes 21\,cm multibeam receiver. Within the 64~MHz bandwidth there are three RRLs: H168$\alpha$, H167$\alpha$ and H166$\alpha$ at 1374.601, 1399.368 and 1424.734\,MHz, respectively.

\begin{figure*}
%\hspace{-0.5cm}
\includegraphics[scale=0.68]{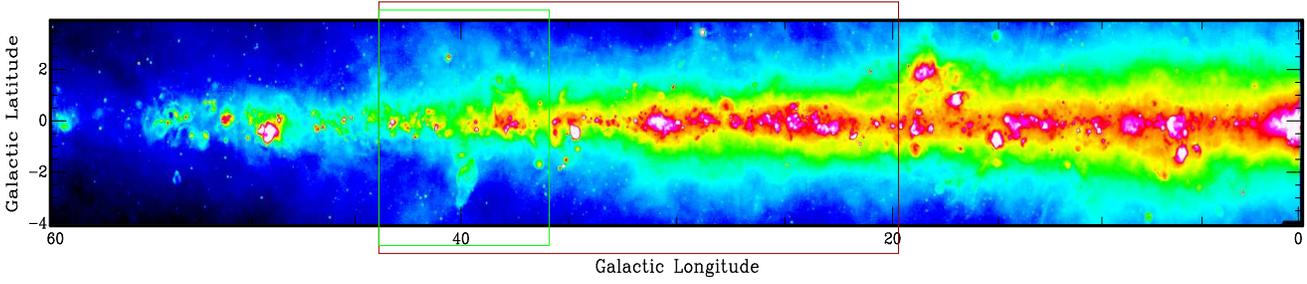}
\caption{The coverage of the $8^{\circ}(\ell) \times 8^{\circ}(b)$ region analysed in \paper~(green) and the larger $24^{\circ}(\ell) \times 8^{\circ}(b)$ area under study in the present work (red). The map is from the 2.7\,GHz survey by \citet{Reich:1990a}.}
\label{fig:bonn1}
\end{figure*}

The HIPASS survey covers the sky south of $\delta=+25\fdg5$. The scans are taken at constant right ascension, in declination strips of 8\degr~separated by 7\,arcmin. The ZOA deep survey is scanned at constant latitude and each 8\degr~longitude scan is separated by 1.4\,arcmin. It covers the Galactic plane accessible from Parkes, $\ell=196^{\circ} - 0^{\circ} - 52^{\circ}$ and $|b| < 5$\degr.
Fig. \ref{fig:bonn1} shows the coverage of the $8^{\circ}(\ell) \times 8^{\circ}(b)$ region centred at $(\ell,b)=(40^{\circ},0^{\circ})$ analysed in \paper~along with the $24^{\circ}(\ell) \times 8^{\circ}(b)$ region analysed in the present work, overlaid on the radio map at 2.7\,GHz from the \citet{Reich:1990a} survey.

The multibeam receiver has 13 beams set on a hexagonal grid with a mean observing beamwidth of 14.4~arcmin. The footprint of the receiver on the sky is $1\fdg7$, thus each scan maps an area of $8^{\circ} \times 1\fdg7$. The multibeam correlator cycle is 5\,s and the scan rate is 1\degr/min. The HIPASS/ZOA frequency coverage is from 1362.5 to 1426.5\,MHz centred on 1394.5\,MHz and is divided into 1024 channels spaced by 62.5\,kHz, or 13.2\kms~for the \hi~line. Therefore an 8\degr~scan is about 100~integrations. The total integration time is 450\,s per beam for the HIPASS survey and 2100\,s per beam for the five times deeper ZOA survey. This results in a typical rms (root mean square) noise of 13\,mJy/beam and 6\,mJy/beam for the HIPASS and ZOA surveys, respectively.

%%%%%%%%%%%%%%%%%%%%%%%%%%%%%%%%%%%%%%%%%%%%%%%%%%%%%%%%%%%%%%%%%%
%%%%%%%%%%%%%%%%%%%%%%%%%%%%%%%%%%%%%%%%%%%%%%%%%%%%%%%%%%%%%%%%%%
\section{THE RRL DATA}
\label{sec:rrldata}
%%%%%%%%%%%%%%%%%%%%%%%%%%%%%%%%%%%%%%%%%%%%%%%%%%%%%%%%%%%%%%%%%%
\subsection{Data Reduction}
\label{sec:datared}
The RRL data reduction pipeline has greatly improved since the first analysis described in \paper.
The previously preferred algorithm for bandpass correction in the {\sc livedata}\footnote{http://www.atnf.csiro.au/computing/software/livedata.html} package \citep{Barnes:2001} was {\sc minmed5}, developed for the treatment of extended emission from High Velocity Clouds (HVCs) \citep{Putman:2002}. The refinement of {\sc minmed5}, the $minmed_{n}$ algorithm, was tested in the new analysis. This method is the running median in a box of size $n$ integrations computed along the length of the scan, for which the bandpass correction is the minimum of those medians. $minmed_{n}$ was found to produce spectral negatives and loss of emission on the Galactic plane, similarly to {\sc minmed5}. The spectral negatives were found to arise from large-scale distortions of the bandpass response by very strong continuum sources, for which the system temperature $T_{\rm sys}$ increases by nearly a factor of two. 
For these reasons, a new algorithm was developed for the exclusive treatment of extended emission from RRLs in {\sc livedata}, called $tsysmin_{n}$. 

\begin{figure*}
\center
\includegraphics[scale=0.4,angle=270]{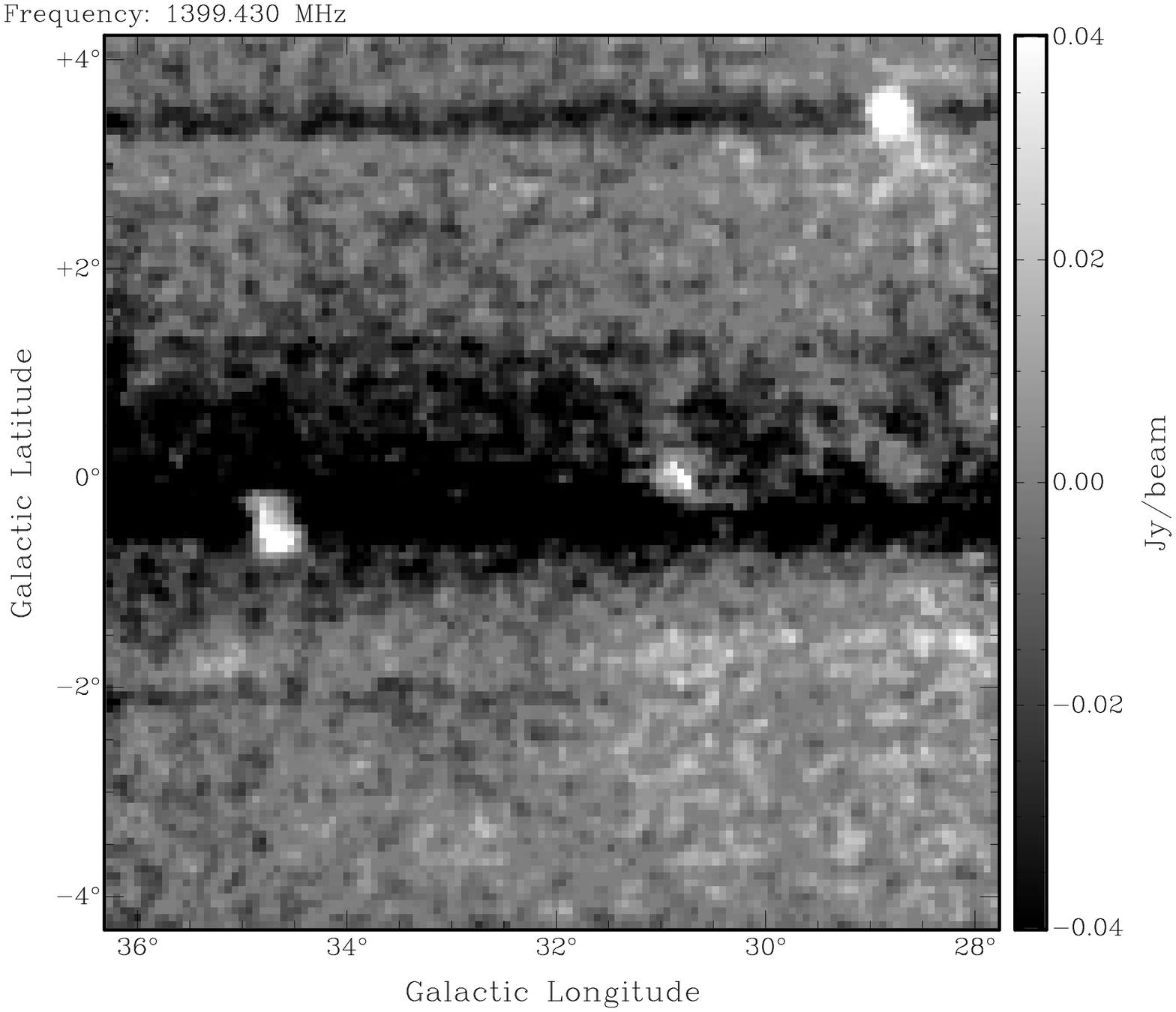}
\includegraphics[scale=0.4,angle=270]{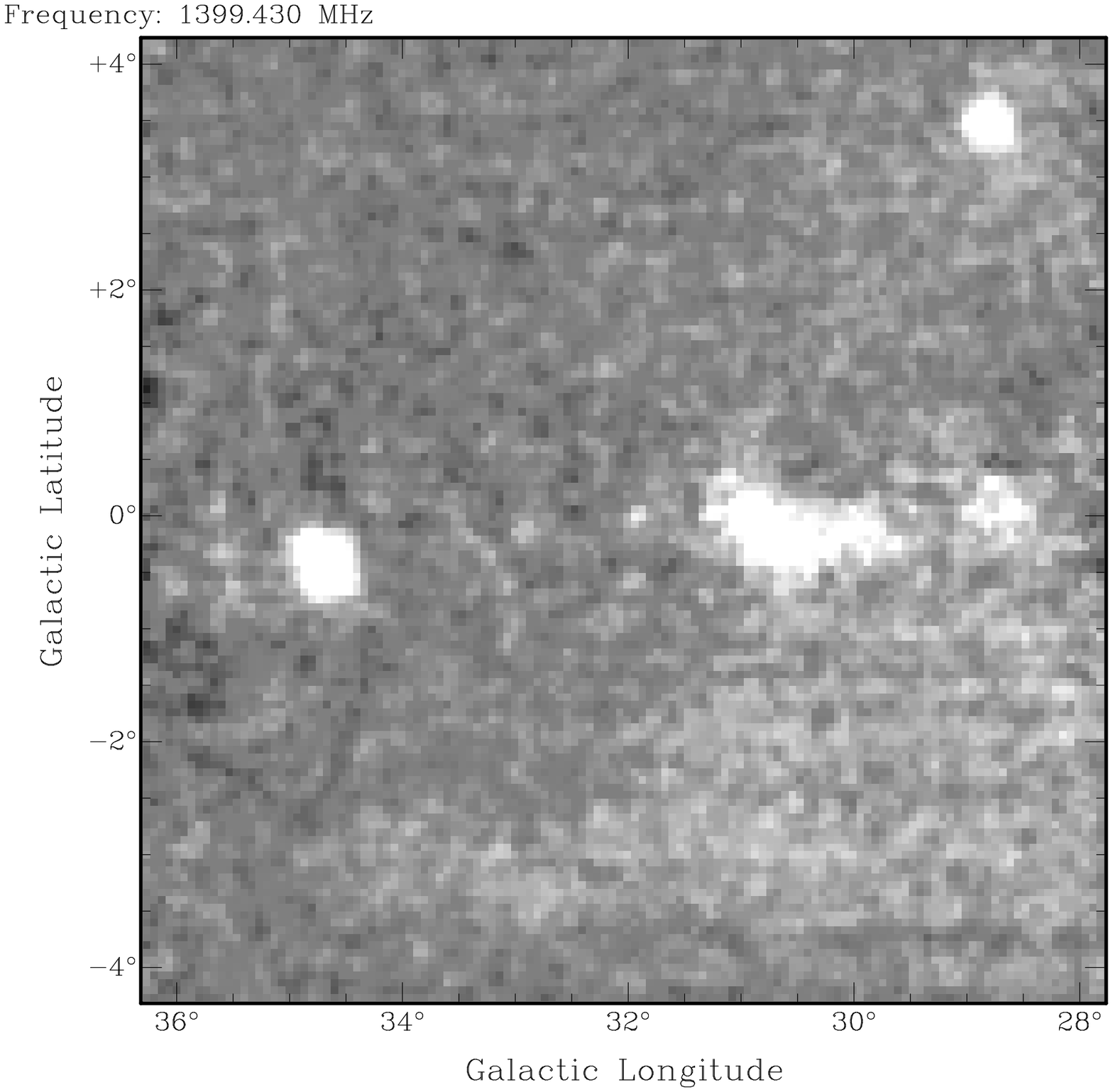}
\hspace{-0.5cm}
\includegraphics[scale=0.34,angle=90]{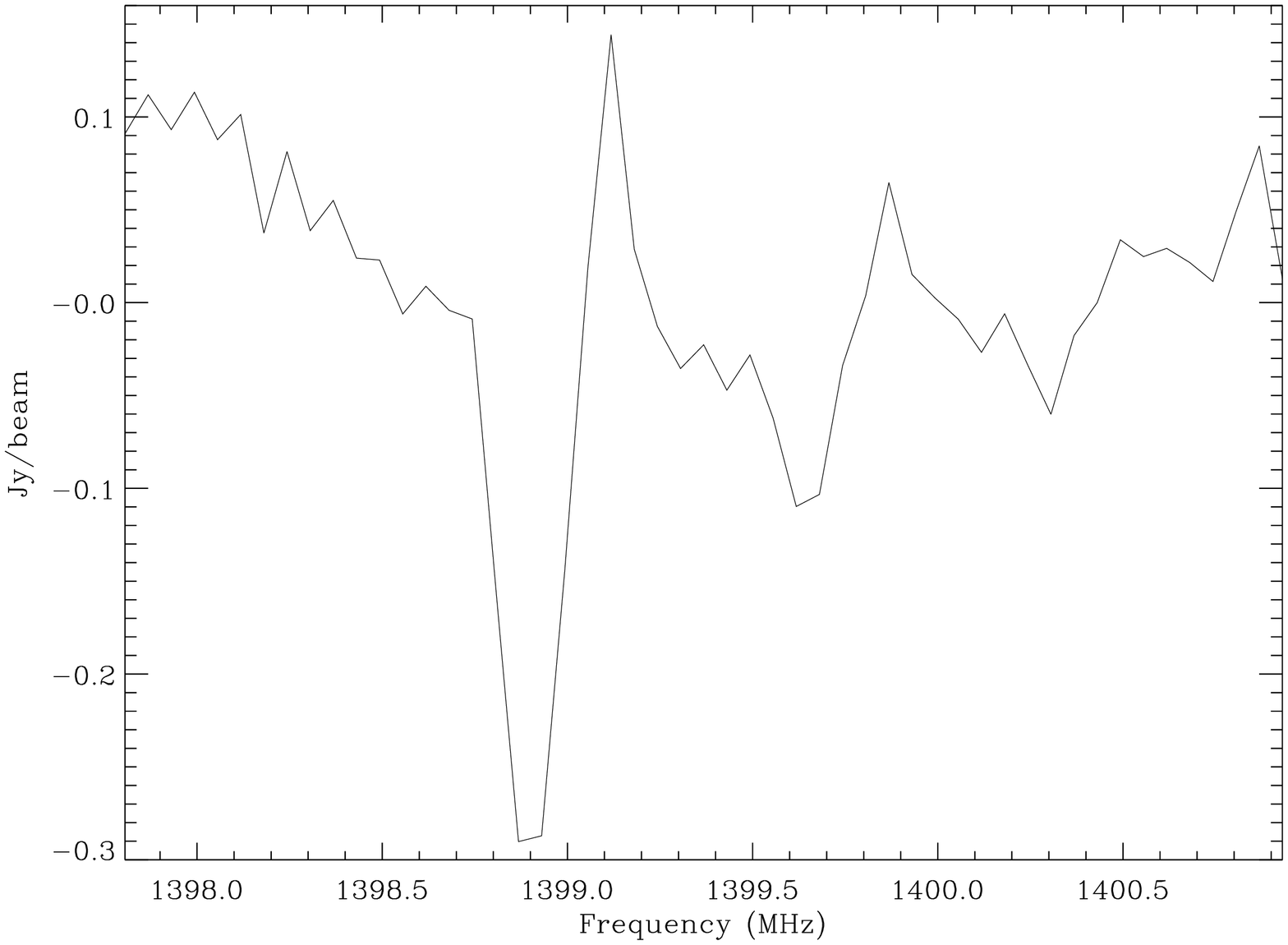}
\includegraphics[scale=0.34,angle=90]{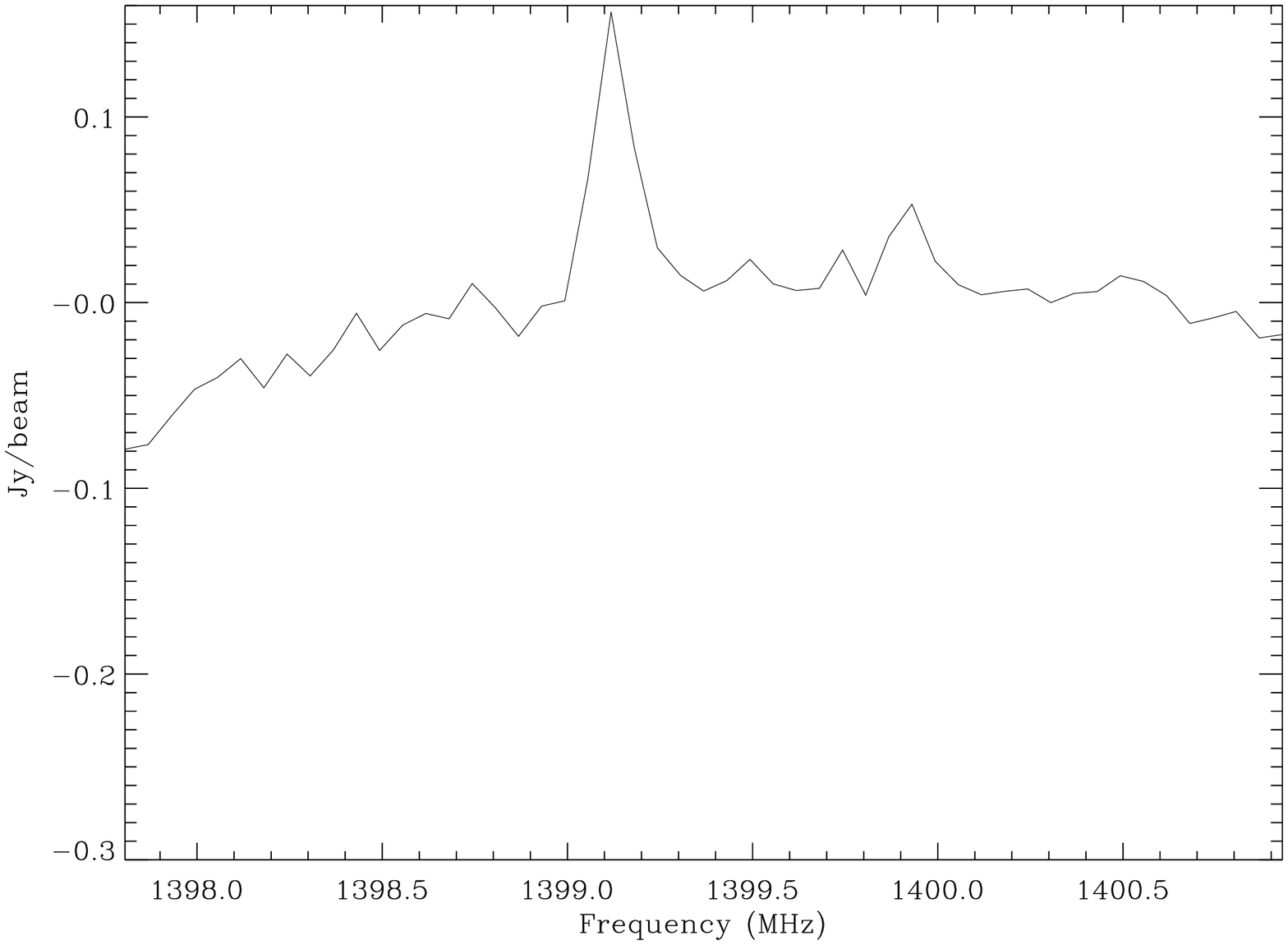}
\caption{(Top) Maps of the ZOA-032 datacube at $V=-13.4$\kms~reduced with $minmed_{10}$~(left) and $tsysmin_{10}$~(right). (Bottom) The corresponding H167$\alpha$ spectra at $(\ell,b)=(35\fdg6,0\fdg0)$.}
\label{fig:tsysmin}
\end{figure*}

$tsysmin_{n}$ locates the $n$ integrations in a scan for which $T_{\rm sys}$ is minimum and, for each spectral channel, takes the median value over these integrations as the bandpass correction. $n=10$ provides a sufficient number of integrations to compute a reliable median without the averaging length being too extended in regions where a large part of the scan may be occupied by extended emission. Fig. \ref{fig:tsysmin} shows a channel map of the ZOA-032 datacube at $V=-13.4$\kms, reduced with $minmed_{10}$ and $tsysmin_{10}$, and H167$\alpha$~spectra towards G35.6+0.0. The $tsysmin$ estimator is used because it is much less susceptible to baseline negatives, renders cleaner spectra with higher peak values and recovers more extended emission on the Galactic plane.

After bandpass correction and calibration the spectra are smoothed using the Tukey filter to suppress the Gibbs ringing caused by the strong Galactic \hi~line, which affects the H166$\alpha$~RRL. The three RRLs are extracted from each spectrum, shifted in velocity and stacked using a weighted mean, where the weights are proportional to the square root of the bandpass response measured at each line's rest frequency. The weights are 0.35, 0.39 and 0.26 for the H168$\alpha$, H167$\alpha$ and H166$\alpha$, respectively.
%Line stacking improves the signal-to-noise by a factor of $\sqrt3$. 

The 3D $\ell-b-$velocity cube for each $8^{\circ} \times 8^{\circ}$ zone is generated from the weighted median of the stacked spectra within a cut-off radius of 6\,arcmin of each pixel, using the package {\sc gridzilla}. The weight for each input spectrum is based on its angular distance from the pixel and its system temperature. The resulting data cube has 4\,arcmin$^{2}$~pixels and a LSR velocity range of $\pm 335$\kms.

\begin{figure*}
\hspace{-0.5cm}
\includegraphics[scale=0.55]{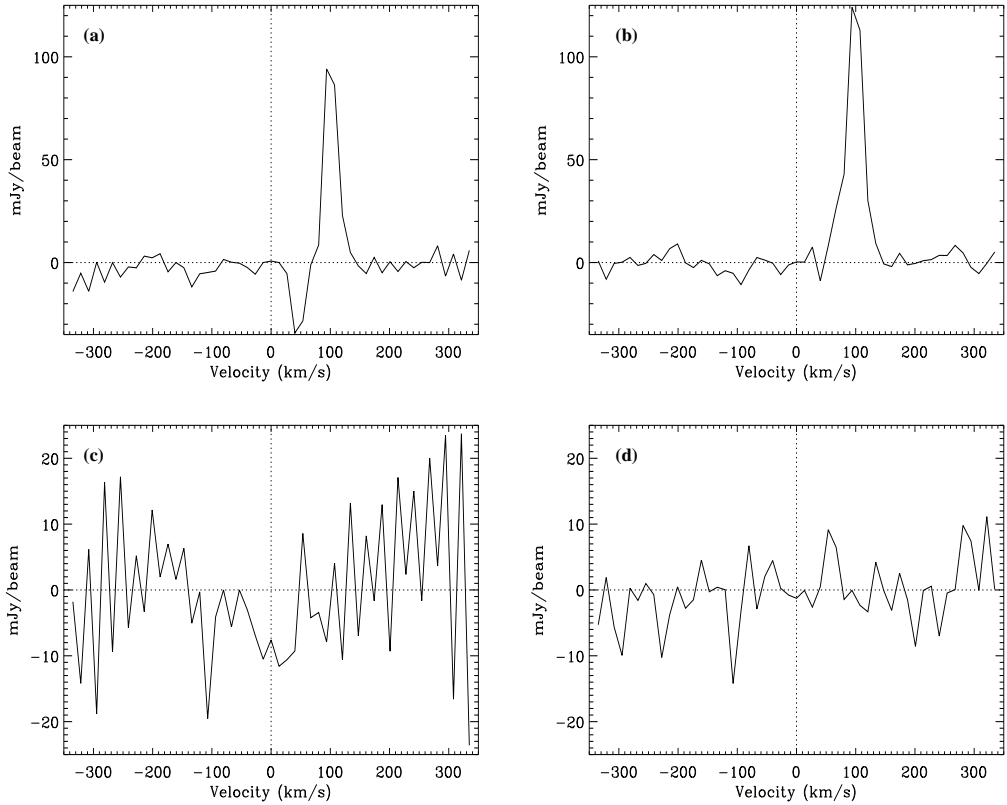}
\caption{ZOA (left) and ZOA+HIPASS spectra after DC level correction (right) at two positions: (top) G33.1-0.1 and (bottom) G28.6-2.1. The improvement after combining both datasets is visible even for spectra away from the Galactic plane.}
\label{fig:rrlcomb}
\end{figure*}

Further improvement of the spectra was achieved by combining HIPASS data with ZOA data. The HIPASS declination scans cross the plane at an angle of $\sim 30$\degr~in the longitude range under study here and therefore yield a real minimum for the bandpass estimation, as opposed to the longitude ZOA scans. We first calculate the DC level between the two surveys by subtracting the ZOA cube from the HIPASS cube, after going through the same reduction pipeline. The DC offset added back to the ZOA spectra is computed as the median of the difference cube at each Galactic latitude and spectral plane. After correction, both datasets are averaged giving 5 times more weight to the ZOA survey since it is 5 times deeper than the HIPASS survey. Finally, spectra in the combined cube are further Tukey smoothed and baseline removed by fitting a polynomial.
Fig. \ref{fig:rrlcomb} shows two examples of spectra improved with the DC level correction and combination of ZOA and HIPASS data. The negative at $V\sim40$\kms~in the ZOA spectrum of G33.1-0.1 is removed after combined with HIPASS. The improvement of the baselines is clear for spectra on and away from the plane, as Fig. \ref{fig:rrlcomb} (d) shows.

%%%%%%%%%%%%%%%%%%%%%%%%%%%%%%%%%%%%%%%%%%%%%%%%%%%%%%%%%%%%%%%%%%
\subsection{Final Data Cube}
\label{sec:finaldata}

The three $8^{\circ} \times 8^{\circ} \times 51$~channel datacubes are combined into a final $24^{\circ} \times 8^{\circ} \times 51$~channel datacube that covers the $\ell$-range 20\degr~to 44\degr, $|b| \le 4$\degr~and $V_{\rm LSR} = \pm 335$\kms. The final spectral resolution after Tukey+Tukey smoothing is degraded from 16\kms~to 20\kms. The typical noise level in each channel is 3.5\,mJy/beam or 3.0\,mK.
The final spatial resolution was measured on the total integrated RRL map using two compact \hii~regions, W49 and W40. The average beam size is 14.8\,arcmin with an uncertainty of 0.6\,arcmin, which is consistent with the intrinsic resolution of 14.4\,arcmin. It is also consistent with simulations, which inject false Gaussian sources into to the data prior to imaging, where, with weighted median gridding, the broadening of the intrinsic 14.4\,arcmin beam is negligible.
The brightness temperature conversion of 0.84\,K/Jy is obtained from the Rayleigh-Jeans relation for a Gaussian FWHM (Full Width at Half Maximum) of 14.4\,arcmin at the frequency of the H167$\alpha$~line. This was checked using well-known bright sources and has an uncertainty of 0.03\,K/Jy. The conversion factor is consistent with previous measurements by \citet{Staveley-Smith:2003} and \citet{Kalberla:2010} that indicate a value of 0.80\,K/Jy.

The total RRL emission map is obtained by integrating the spectra in the appropriate velocity range, to include the different contributions from the Local ($V=0$\kms), Sagittarius ($V=40$\kms) and Scutum ($V=80$\kms) arms. The integral varies between $-20$ and $150$\kms~in order to account for any velocity spread in within the arms. The rms of the $24^{\circ} \times 8^{\circ}$ RRL integrated map is 0.05\,K\kms, and the maximum and minimum are $-0.2$ and $26.7$\,K\kms, respectively.

The data are calibrated in the main beam scale, appropriate for point sources, rather than on the full-beam which includes the near sidelobes \citep{Reich:1988}. However due to the imaging process the flux for point sources is lowered in the final cubes. This depends on the cut-off radius used in {\sc gridzilla} since the final value for a given pixel is the median of all the data that are within a certain distance from it. Simulations show that for a cut-off radius of 6\,arcmin the peak fluxes of point sources are underestimated by 20 per cent. Further investigation using the datacubes reduced with smaller cut-off radii, for which this effect is increasingly small, show that it only affects the two most compact \hii~regions W49 and W40 which are about 5\,arcmin in size. The 6\,arcmin cut-off radius is used because it provides an adequate flux scale with an uncertainty of approximately 10 per cent for extended sources, in a low noise cube.

The total power map from the same survey is available for the region under study (Calabretta et al., in prep.) and is shown in Fig. \ref{fig:maps}. It is reduced using ZOA and HIPASS data similarly to the RRL analysis, in order to recover the zero levels on the Galactic plane and gridded with the same parameters.

%%%%%%%%%%%%%%%%%%%%%%%%%%%%%%%%%%%%%%%%%%%%%%%%%%%%%%%%%%%%%%%%%%
\subsection{Helium and Carbon RRLs}
\label{sec:finaldata}

\begin{figure}
\hspace{-0.6cm}
\includegraphics[scale=0.35,angle=90]{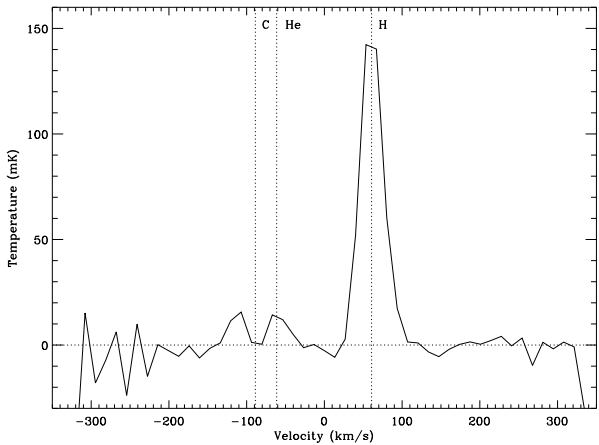}
\caption{Spectrum for the \hii~region W47 $(\ell,b)=(37\fdg8,-0\fdg3)$ showing the H, He and C RRLs. The vertical dotted lines indicate the rest velocity of each line. The C RRL is amplified by maser emission by the foreground gas and shifted to lower velocities.}
\label{fig:helium}
\end{figure}

Helium and Carbon RRLs are observed in the HIPASS/ZOA spectra. The expected velocity separation of 122.1\kms~and 149.5\kms~between the hydrogen RRL and the helium and carbon lines, respectively, allow their identification. Due to the spectral resolution of 20\kms, these two RRLs are sometimes blended. When the carbon RRL is amplified by maser emission by foreground gas, it shifts towards lower velocities and the two can be separated. An example is shown in Fig. \ref{fig:helium}, for the \hii~region W47.

Further investigation will provide information on the He/C RRLs from the diffuse ionised gas in the inner Galaxy. This will be presented in a future work.

%%%%%%%%%%%%%%%%%%%%%%%%%%%%%%%%%%%%%%%%%%%%%%%%%%%%%%%%%%%%%%%%%%
%%%%%%%%%%%%%%%%%%%%%%%%%%%%%%%%%%%%%%%%%%%%%%%%%%%%%%%%%%%%%%%%%%
\section{CONVERSION TO A FREE-FREE TEMPLATE}
\label{sec:fftemp}

The RRL emission from a diffuse ionised hydrogen gas can be expressed
in terms of the integral over spectral line temperature as
\begin{equation}
\int T_{\rm L} d \nu = 1.92 \times 10^{3}  T_{\rm e}^{-1.5}  {\rm EM} 
\label{eq:1}
\end{equation}
where $\nu$ is the frequency (kHz), $T_{\rm e}$ is the electron temperature (K) and the
emission measure EM is in cm$^{-6}$pc.  The corresponding continuum
emission brightness temperature is
\begin{equation}
T_{\rm b} = 8.235 \times 10^{-2} a(T_{\rm e})  T_{\rm e}^{-0.35} \nu^{-2.1}_{\rm GHz}
 (1+0.08)  {\rm EM}
\label{eq:2}
\end{equation}
where $a(T_{\rm e})$ is a slowly varying function of temperature and $\nu$
is the frequency in GHz \citep{Mezger&Henderson:1967}.  The (1+0.08) term represents the additional
contribution to $T_{\rm b}$ from helium.  These equations lead to an
expression for the ratio of the line integral to the continuum
\begin{equation}
\frac{\int T_{\rm L} dV}{T_{\rm b}} = 6.985 \times 10^{3} \frac{1}{a(T_{\rm e})}
 \frac{1}{n(\rm He)/(1 + n(\rm H))}  T_{\rm e}^{-1.15}  \nu^{1.1}_{\rm GHz} 
\label{eq:3}
\end{equation}

where $V$ is in \kms~\citep{Rohlfs:2004}.

%%%%%%%%%%%%%%%%%%%%%%%%%%%%%%%%%%%%%%%%%%%%%%%%%%%%%%%%%%%%%%%%%%
\subsection{The $T_{\rm e}$ of the thermal emission from $\ell=20$\degr~to $44$\degr}
\label{sec:te}
In order to convert the RRL integral into a free-free brightness temperature, a value for the electron temperature of the ionised gas is needed. The variation of $T_{\rm e}$ in discrete \hii~regions with longitude, or Galactic radius $R_{\rm G}$, is well-known \citep{Shaver:1983,Paladini:2004}. It is caused by the greater cooling of the \hii~regions closer to the Galactic centre due to a higher metal content in the inner Galaxy. \citet{Paladini:2004} uses all the published data for 404 \hii~regions with reliable Galactocentric distances to derive
\begin{equation}
T_{\rm e}[{\rm K}] = (4166 \pm 124) + (314 \pm 20) R_{\rm G} [{\rm kpc}].
\label{eq:paladini}
\end{equation} 
It follows that the mean $T_{\rm e}$ at the Galactic radius of the Local arm, 8\,kpc, is 6680\,K and decreases to 5740\,K at 5.0\,kpc, the radius of the Scutum arm.

The line-to continuum ratio given by equation (\ref{eq:3}) is one of the most reliable methods to determine the electron temperature of an \hii~region and this survey has the advantage of providing both the line and continuum data at the same resolution, after going through the same reduction pipeline. However, $T_{\rm b}$ in equation (\ref{eq:3}) must correspond to the thermal component only, thus an estimate of the synchrotron emission within the beam must be removed from the total continuum. 

\begin{figure*}
\hspace{-0.5cm}
\includegraphics[scale=0.55]{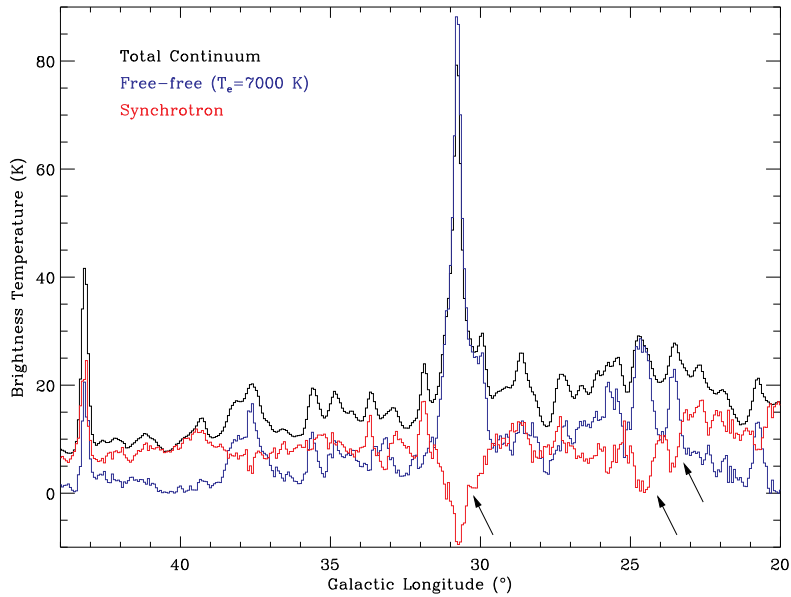}
\caption{Comparison between the total continuum, free-free estimated from the RRLs with $T_{\rm e}=7000$\,K and the synchrotron, versus longitude at $b=0$\degr. The arrows show three regions of decrease in the derived synchrotron emission coincident in shape and position with \hii~regions.}
\label{fig:te1}
\end{figure*}

Fig. \ref{fig:te1} shows the distribution of the total continuum, the free-free emission estimated from the RRLs with the Local $T_{\rm e}$ of 7000\,K and the difference, that corresponds to the synchrotron emission, at $b=0$\degr. The arrows indicate three regions of decrease in the derived synchrotron emission coincident in shape and position with \hii~regions. This is due to an overestimation of the free-free caused by an overestimation of $T_{\rm e}$. 
The opposite occurs for example for the \hii~region W40, where an increase in the synchrotron distribution is correlated with the free-free emission. This indicates that the electron temperature of this local \hii~region is higher than 7000\,K.
Fig. \ref{fig:te1} also shows that the synchrotron longitude distribution is relatively smooth. 
We use this fact to estimate the synchrotron emission underlying \hii~regions situated away from any strong synchrotron emitting source.
The synchrotron background is taken as the average of its longitude distribution across an \hii~region and either side of it.

We selected 16 \hii~regions in the area under study that have well-defined longitude profiles and spectra with only one velocity component, for which $T_{\rm e}$ is obtained from equation (\ref{eq:3}). Using the longitude profiles taken at the central latitude of each \hii~region, we estimate the object's longitude extent from its free-free distribution. The synchrotron background, $T_{sync}$, is the average of the synchrotron distribution either side of the \hii~region. The thermal continuum $T_{\rm b}$ is calculated from the difference of the total $T_{\rm C}$ and the synchrotron temperatures. These are measured at the centre of the object, along with the line integral $\int T_{\rm L} \Delta V$ in equation (\ref{eq:3}).
The uncertainties on these quantities are the rms of the corresponding maps measured away from the Galactic plane, typically at $|b|=3^{\circ}-4^{\circ}$. Table \ref{table:te} lists the \hii~regions with the corresponding RRL velocities and Galactocentric radii, along with the derived $T_{\rm e}$. The Galactocentric distance is calculated using the \citet{Fich:1989} rotation curve, with $R_{0}=8.5$\,kpc, assuming that the object moves around the Galactic centre in a circular orbit with radial velocity $V$.

\begin{table*}
\centering\small
\caption{List of \hii~regions selected for the study of the electron temperature variation with Galactocentric distance. $R_{\rm G}$, in column 3 is calculated using the RRL central velocity and the \citet{Fich:1989} rotation curve with $R_{0}=8.5$\,kpc. The velocities and corresponding uncertainties, in column 2, are the result of a Gaussian fit. $T_{\rm C}$, $T_{sync}$, $T_{ff}$, $\int T_{\rm L} \Delta V$ are obtained from the longitude profiles using the method described in the text; the uncertainties, taken as the rms measured on the maps away from the Galactic plane, are 0.6\,K, 0.6\,K, 0.2\,K and 0.5\,K\kms, respectively. $T_{\rm e}$ is calculated using equation (\ref{eq:3}). The last column identifies each \hii~region by its commonly used name \citep{Paladini:2003}.
\label{table:te}}
\begin{tabular}{lcccccccc}
\\
\hline
\hline
{\small \hii~Region} & $V$ & $R_{\rm G}$ & $T_{\rm C}$ & $T_{sync}$ & $T_{ff}$ & $\int T_{\rm L} \Delta V$ & $T_{\rm e}$ & Name\\ 
& {\small(\kms)} & {\small(kpc)} & {\small(K)} & {\small(K)}& {\small(K)}& {\small(K\kms)}& {\small(K)} &  \\
\hline
{\small G40.5+2.5} & $25.9 \pm 0.7$ & 7.1 & 9.5 & 5.0 & 4.5 & 1.3 & $8360 \pm 1590$ & W45\\
{\small G37.8-0.1} & $59 \pm 1$     & 5.9 & 21.1 & 6.4 & 14.7 & 6.0 & $6180 \pm 340$ & W47\\
{\small G37.7-0.2} & $60 \pm 1$     & 5.9 & 22.3 & 7.3 & 15.0 & 5.8 & $6520 \pm 350$ & \\
{\small G36.3-1.7} & $62.2 \pm 0.5$ & 5.7 & 7.0 & 3.8 & 3.2 & 1.3 & $6210 \pm 1480$ & RWC179\\
{\small G35.6+0.0} & $53.4 \pm 0.3$ & 6.0 & 19.4 & 8.8 & 10.6 & 4.0 & $6650 \pm 410$ & \\
{\small G33.1-0.1} & $98.2 \pm 0.4$ & 4.7 & 15.7 & 8.6 & 7.1 & 3.3 & $5560 \pm 530$ & Ke78\\
{\small G30.9-0.1} & $97.5 \pm 0.6$ & 4.6 & 61.0 & 8.9 & 52.1 & 23.7 & $5660 \pm 190$ & W43\\
{\small G30.7+0.0} & $95.6 \pm 0.6$ & 4.6 & 77.3 & 10.4 & 66.9 & 30.9 & $5580 \pm 170$& W43\\
{\small G30.0-0.1} & $97.3 \pm 0.4$ & 4.5 & 61.0 & 8.9 & 52.1 & 23.7 & $5780 \pm 280$&\\
{\small G28.8+3.5} & $1.5 \pm 0.4$  & 8.3 & 24.7 & 6.0 & 18.7 & 5.1 & $8770 \pm 560$ & W40\\
{\small G27.1+0.0} & $92.9 \pm 0.5$ & 4.4 & 20.9 & 8.6 & 12.3 & 4.8 & $6500 \pm 440$ & 3C387\\
{\small G25.8+0.2} & $107.6 \pm 0.4$& 4.0 & 27.7 & 10.0 & 17.6 & 7.4 & $6060 \pm 290$ & \\
{\small G24.8+0.1} & $108.0 \pm 0.4$& 3.9 & 28.9 & 8.4 & 20.5 & 9.1 & $5790 \pm 240$& W42\\
{\small G23.5+0.0} & $92.2 \pm 0.4$ & 4.1 & 27.4 & 12.0 & 15.4 & 7.4 & $5400 \pm 280$ & \\ 
{\small G22.9-0.3} & $74.8 \pm 0.4$ & 4.5 & 33.7 & 12.5 & 21.2 & 9.2 & $5870 \pm 220$ & W41\\
{\small G20.7-0.1} & $53.6 \pm 0.3$ & 5.0 & 23.2 & 10.3 & 12.9 & 6.3 & $5340 \pm 290$ & Ke68\\
\hline
\end{tabular}
\end{table*}

\begin{figure}
\hspace{-0.5cm}
\includegraphics[scale=0.5]{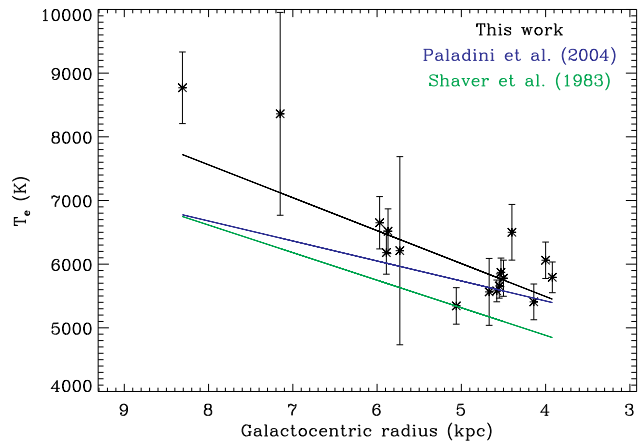}
\caption{The $T_{\rm e}-R_{\rm G}$ relationship derived from the results in Table \ref{table:te}. The black line gives the best linear fit to the data points (equation (\ref{eq:te})) and the blue and green lines are the \citet{Paladini:2004} and \citet{Shaver:1983} results, respectively.}
\label{fig:tefit}
\end{figure}

The electron temperatures obtained for the 16 \hii~regions are shown in Fig. \ref{fig:tefit} as a function of the Galactocentric distance. The best linear fit to the data gives:
\begin{equation}
T_{\rm e} = (3423 \pm 472) + (517 \pm 99)R_{\rm G}
\label{eq:te}
\end{equation}
where the uncertainties on $T_{\rm e}$ are taken into account by the least-square fitting procedure.
As shown in Fig. \ref{fig:tefit} this result is in general agreement with the \citet{Paladini:2004} and \citet{Shaver:1983} results with larger samples. The difference between the three lines is mainly caused by the lack of data points beyond the solar radius in the present sample and the two higher $T_{\rm e}$ values from the local \hii~regions W40 and W45. 
There is an intrinsic scatter at any given $R_{\rm G}$ in all studies due to the different properties of each \hii~region. 

%%%%%%%%%%%%%%%%%%%%%%%%%%%%%%%%%%%%%%%%%%%%%%%%%%%%%%%%%%%%%%%%%%
\subsection{The total free-free brightness temperature}
\label{sec:ff}

We are now in a position to convert the RRL integral into free-free brightness temperature at each pixel using the derived $T_{\rm e}-R_{\rm G}$ relationship in equation (\ref{eq:te}). The Galactocentric distance is calculated using the velocity at peak of each spectra. The maximum and minimum of the derived $T_{\rm e}$ map for this region are 4900\,K and 8590\,K, respectively. The average $T_{\rm e}$ at $b=0$\degr~over the longitude range studied is $6080 \pm 970$\,K. The fact that we are ignoring a possible second weaker velocity component does not affect the resulting $T_{\rm e}$ by more than its uncertainty.
%, also taking into account the uncertainties on the estimation of $V$ and $R_{\rm G}$.

\begin{figure*}
\includegraphics[scale=0.85]{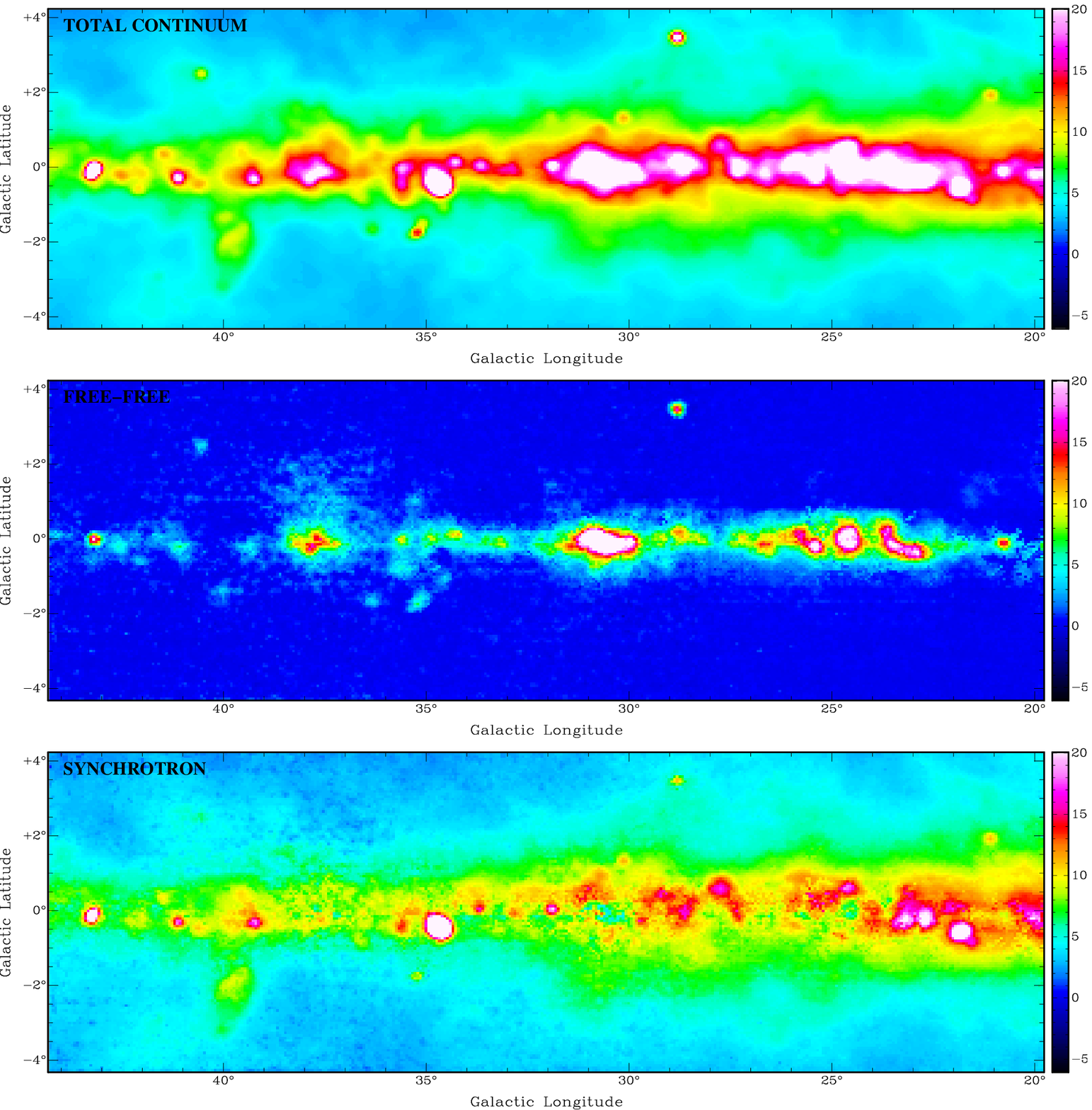}
\caption{Maps of the total continuum, free-free and synchrotron emission at 1.4\,GHz and 14.8\,arcmin resolution. The free-free is estimated from the RRL integral using the $T_{\rm e}-R_{\rm G}$ relationship from equation (\ref{eq:te}). The synchrotron is the difference between the total continuum and the free-free emission and shows a narrow diffuse emission confined to the plane. The colour scale is linear and in units of brightness temperature (K).}
\label{fig:maps}
\end{figure*}

The final free-free map is shown in Fig. \ref{fig:maps}, along with total and synchrotron emission maps for the same region. The synchrotron map will be discussed in Section \ref{sec:synctemp}.
The free-free map shows the thermal emission emanating from particular objects, mostly lying close to the Galactic plane, whereas the diffuse component of the total continuum is mainly associated with the non-thermal emission. While there is evidence of incomplete separation, e.g. in the W40 \hii~region, most of the compact sources are well separated as either \hii~regions or SNRs, and that, along with the diffuse emission, provides a visual indication of the reliability of the method.

%%%%%%%%%%%%%%%%%%%%%%%%%%%%%%%%%%%%%%%%%%%%%%%%%%%%%%%%%%%%%%%%%%
\subsection{Catalogue of \hii~regions}
\label{sec:hiicat}

A catalogue of \hii~regions from the present free-free map was created using the source extractor program SExtractor \citep{Bertin:1996}. 
The main parameters to adjust in SExtractor are the background mesh size, detection threshold and minimum area which are set to 16, $2\sigma$ and 2 pixels, respectively. The filter function used is a Mexican hat filter of width equal to 3\,pixels. The mesh size is equivalent to $\simeq 1$\degr~and the width of the filter function is similar to the 14.8\,arcmin beam. The estimated background rms by SExtractor is 0.26\,Jy/beam, so the threshold limit is 0.53\,Jy/beam. The contrast parameter, which controls the deblending of multiple sources, is set to 0.005. These parameters were tuned to recover objects on the Galactic plane with sizes consistent with those from higher resolution surveys, such has the 4.3\,arcmin resolution survey at 2.7\,GHz by \citet{Reich:1990a}.

Table \ref{tab:sextparam_hii} lists the 57 \hii~regions extracted from the free-free map. After the source is detected in the filtered map, it is fitted for an elliptical shape, using the second order moments, whose parameters $a$ and $b$ correspond to the ellipse major and minor axis, respectively. The position angle (P.A.) listed is the angle between the semi-major axis of the source and the longitude axis, measured counter-clockwise. The peak flux of the source is measured directly from the input map and subtracted by the background. The uncertainty on the peak flux is estimated using the rms uncertainties on the free-free and background maps. Since SExtractor does not provide flux densities from Gaussian fitting, we use the peak flux and the observed size ($a \times b$) to calculate it assuming a Gaussian profile for the source:
\begin{equation}
S = S_{\rm P} \frac{a \times b}{\mbox{fwhm}^{2}}
\label{eq:sextflux}
\end{equation}
where $S_{\rm P}$ is the peak flux in Jy/beam and fwhm is the beam FWHM of 14.8\,arcmin. The uncertainty on $S$ is estimated using the uncertainties on $S_{\rm P}$ and on $a$ and $b$, which are given by SExtractor. The maximum uncertainty on the observed size is 4\,arcmin. A source of 5\,arcmin in size is observed by the 14.8\,arcmin beam with a diameter of 15.6\,arcmin. This small broadening of the beamwidth is difficult to measure, especially for a weak object within the noise. For the objects whose apparent area is smaller than the beam area, the flux density is the same as the peak,  $S = S_{\rm P}$. In such cases, the P.A. has a large uncertainty. The flag parameter in Table \ref{tab:sextparam_hii} indicates if: 1 the object has bright neighbours; 2 the source was originally blended with another one; 3 that cases 1 and 2 apply. It shows that deblending occurs mainly in the \hii~complexes W47, W43 and W42.

\begin{table*}
\centering\small
\caption{\hii~regions in the region $\ell=44$\degr~to $20$\degr, extracted from the free-free map using SExtractor. Column 1 numbers each object; columns 2 and 3 are the coordinates; columns 4 and 5 give the angular size of the source, thus $a$ and $b$ deconvolved with the 14.8\,arcmin beam, which is $p$ when the object is less than a few ($\sim 5$)\,arcmin; column 6 is the angle measured counter-clockwise between the semi-major axis of the source and the longitude axis; columns 7 and 8 give the peak and flux density at 1.4\,GHz; column 9 gives the internal flags generated by SExtractor (see text); column 10 gives the RRL central velocity from a one or two component Gaussian fit; column 11 identifies a source by its commonly used name as in the \citet{Paladini:2003} catalogue.\label{tab:sextparam_hii}}
\begin{tabular}{ccccccccccc}
\hline \hline 
{\small Number} & $\ell$ & $b$ & $\theta_{a}$ & $\theta_{b}$ & {\small P.A.} & {\small Peak} & {\small Flux} & {\small Flag} & {\small $V$} &{\small Name}\\ 
& ($^{\circ}$) & ($^{\circ}$) & ($^{\prime}$) & ($^{\prime}$) & ($^{\circ}$) & {\small (Jy/beam)} & {\small (Jy)}& {\small \kms} && \\ 
\hline 
$ 1$&$43.18$&$ 0.00$&$ 10.4$&$ 6.3$&$-45$&$ 25.6\pm 3.6$&$ 41.2\pm 6.7$&$ 0$ & $9.4$&{\small W49A}\\ 
$ 2$&$42.54$&$-0.21$&$ 9.7 $&$ p $&$-76$&$ 6.0\pm 1.3$&$ 6.8\pm1.8$&$ 0$ & $20.4,67.3$& \\ 
$ 3$&$42.06$&$-0.58$&$ 1.7 $&$ p $&$ -9$&$ 3.7\pm 0.9$&$ 3.7\pm 0.9$&$ 0$ & $66.6$& \\ 
$ 4$&$41.63$&$ 0.10$&$ 4.9 $&$ p $&$-19$&$ 4.7\pm 0.9$&$ 4.7\pm 0.9$&$ 0$ & $14.9,53.6$&\\ 
$ 5$&$41.08$&$-0.19$&$ 21.7$&$ p $&$ 11$&$ 8.0\pm 1.2$&$ 10.1\pm 2.0$&$ 0$ & $58.8$&\\ 
$ 6$&$40.54$&$ 2.51$&$ 6.1 $&$ p $&$ 68$&$ 4.5\pm 1.0$&$ 4.5\pm 1.0$&$ 0$ & $26.2$&{\small W45}\\ 
$ 7$&$40.00$&$-1.37$&$ p $&$ p $&$ 0$&$ 4.1\pm 1.2$&$ 4.1\pm 1.2$&$ 0$ & $47.7$&{\small S74}\\ 
$ 8$&$39.38$&$-0.17$&$ 6.8 $&$ p $&$ 33$&$ 5.0\pm 1.3$&$ 5.0\pm 1.3$&$ 2$ & $19.4,60.9$&{\small NRAO591}\\ 
$ 9$&$38.23$&$-0.14$&$ 12.2$&$ p $&$ 19$&$ 12.4\pm 2.9$&$ 12.4\pm 2.9$&$ 1$ &$56.1,86.9 $&\\
$10$&$37.84$&$-0.25$&$ 17.1$&$ 1.7$&$ 47$&$ 17.8\pm 3.9$&$ 27.3\pm 6.5$&$ 3$ & $60.7$&{\small W47}\\ 
$11$&$37.60$&$-0.01$&$ 32.2$&$ p $&$-43$&$ 16.6\pm 3.3$&$ 32.0\pm 7.3$&$ 2$ & $52.2,86.7$&{\small W47}\\
$12$&$37.21$&$-0.12$&$ 5.5 $&$ p $&$ 24$&$ 12.0\pm 2.5$&$ 12.0\pm 2.5$&$ 3$ &$42.5,80.2$&\\ 
$13$&$36.91$&$-0.16$&$ p $&$ p $&$ 46$&$ 6.2\pm 2.0$&$ 6.2\pm 2.0$&$ 0$ &$40.9,80.4$ &\\ 
$14$&$36.50$&$-0.20$&$ p $&$ p $&$ 0$&$ 3.3\pm 1.8$&$ 3.3\pm 1.8$&$ 0$ & $46.6,73.9$&\\ 
$15$&$36.34$&$-1.67$&$ 2.4 $&$ p $&$ -9$&$ 4.3\pm 1.1$&$ 4.3\pm 1.1$&$ 0$ & $62.1$&{\small RCW179}\\
$16$&$35.65$&$-0.81$&$ 5.2 $&$ p $&$ 45$&$ 5.0\pm 1.7$&$ 5.0\pm 1.7$&$ 2$ &$58.8$&\\ 
$17$&$35.59$&$-0.01$&$ 10.1$&$ 7.7$&$ 86$&$ 10.5\pm 2.4$&$ 14.4\pm 3.6$&$ 0$ &$53.4 $&\\ 
$18$&$35.27$&$ 1.03$&$ p $&$ p $&$ 0$&$ 3.8\pm 1.2$&$ 3.8\pm 1.2$&$ 0$ & $76.0$&\\ 
$19$&$35.19$&$-1.69$&$ 27.4$&$ p $&$ 53$&$ 7.9\pm 1.4$&$ 13.7\pm 3.0$&$ 0$ & $45.3$&{\small W48}\\ 
$20$&$34.88$&$ 0.00$&$ 15.9$&$ 11.6$&$ 69$&$ 9.2\pm 2.5$&$ 17.1\pm 5.0$&$ 0$ & $53.3$&\\ 
$21$&$34.74$&$-0.77$&$ p $&$ p $&$ 32$&$ 3.2\pm 1.7$&$ 3.2\pm 1.7$&$ 0$ & $47.3$&{\small W44}\\
$22$&$34.52$&$-1.05$&$ p $&$ p $&$ 47$&$ 3.8\pm 1.2$&$ 3.8\pm 1.2$&$ 0$ &$48.1$ &\\ 
$23$&$34.27$&$ 0.13$&$ 19.9$&$ p $&$ 2$&$ 12.3\pm 3.0$&$ 16.5\pm 4.4$&$ 1$ & $57.2$&{\small W44}\\ 
$24$&$33.57$&$-0.06$&$ 21.2$&$ p $&$ 51$&$ 7.7\pm 2.4$&$ 8.2\pm 2.9$&$ 0$ & $58.9,99.7$&\\
$25$&$33.14$&$-0.10$&$ 16.5$&$ 7.9$&$-45$&$ 9.6\pm 2.4$&$ 16.4\pm 4.4$&$ 0$ & $98.3$&{\small Ke78}\\ 
$26$&$32.22$&$-0.17$&$ 12.9$&$ p $&$-15$&$ 7.7\pm 2.3$&$ 7.7\pm 2.3$&$ 1$ & $39.9,88.0$&\\ 
$27$&$32.17$&$ 0.10$&$ 2.4 $&$ p $&$ 70$&$ 6.6\pm 2.0$&$ 6.6\pm 2.0$&$ 1$ & $44.1,90.6$&\\
$28$&$31.93$&$-0.37$&$ p $&$ p $&$ 2$&$ 8.4\pm 2.5$&$ 8.4\pm 2.5$&$ 1$ & $42.9,88.7$&\\ 
$29$&$31.90$&$ 1.47$&$ p $&$ p $&$ 0$&$ 2.2\pm 0.6$&$ 2.2\pm 0.6$&$ 0$ & $56.7$&{\small RCW177}\\ 
$30$&$31.49$&$-0.21$&$ 6.1 $&$ p $&$ -9$&$ 15.1\pm 3.6$&$ 15.1\pm 3.6$&$ 3$ &$39.0,94.7$ &\\
$31$&$30.86$&$-0.74$&$ 4.9 $&$ p $&$ 12$&$ 8.4\pm 2.8$&$ 8.4\pm 2.8$&$ 0$ & $94.8$&\\ 
$32$&$30.81$&$-0.02$&$ 29.1$&$ 12.2$&$-11$&$ 82.4\pm 16.3$&$ 236.3\pm 50.6$&$ 1$ & $97.6$&{\small W43,M51,Ke76}\\
$33$&$30.27$&$-0.19$&$ 31.9$&$ p $&$ 26$&$ 32.4\pm 6.4$&$ 66.0\pm 14.7$&$ 3$ &$ 99.8$&\\ 
$34$&$29.93$&$-0.05$&$ 18.2$&$ 10.6$&$ 83$&$ 24.2\pm 4.8$&$ 47.5\pm 10.3$&$ 2$ &$ 97.2$&\\ 
$35$&$29.21$&$ 0.08$&$ 14.0$&$ p $&$-29$&$ 10.0\pm 3.2$&$ 10.1\pm 3.6$&$ 0$ &$ 56.9,96.1$&\\
$36$&$29.09$&$-0.65$&$ p $&$ p $&$-73$&$ 4.1\pm 2.4$&$ 4.1\pm 2.4$&$ 0$ & $50.0$&{\small RCW175}\\ 
$37$&$28.80$&$ 3.49$&$ 10.9$&$ 5.8$&$-52$&$ 19.0\pm 3.0$&$ 30.6\pm 5.4$&$ 0$ & $1.5$&{\small W40}\\
$38$&$28.74$&$ 0.17$&$ 24.9$&$ 11.3$&$-43$&$ 13.1\pm 3.7$&$ 32.0\pm 9.6$&$ 1$ &$ 97.2$&\\ 
$39$&$28.18$&$-0.02$&$ 29.0$&$ p $&$-34$&$ 10.1\pm 3.1$&$ 17.9\pm 6.0$&$ 1$ & $94.9$&\\ 
$40$&$27.41$&$-0.13$&$ 6.8 $&$ p $&$-18$&$ 7.8\pm 3.0$&$ 7.8\pm 3.0$&$ 3$ & $92.6$&{\small 3C387}\\
$41$&$27.17$&$ 0.03$&$ 10.6$&$ p $&$-75$&$ 13.0\pm 4.5$&$ 10.7\pm 3.3$&$ 2$ & $93.7$&{\small 3C387}\\ 
$42$&$26.66$&$-0.14$&$ 21.7$&$ p $&$ 2$&$ 14.9\pm 3.9$&$ 22.7\pm 4.5$&$ 1$ &$73.0,101.7$ &\\ 
$43$&$26.13$&$-0.07$&$ p $&$ p $&$ 0$&$ 10.8\pm 4.3$&$ 10.8\pm 4.3$&$ 3$ & $103.2$&\\ 
$44$&$25.77$&$ 0.14$&$ 25.0$&$ 9.9$&$-78$&$ 16.5\pm 5.5$&$ 39.1\pm 13.4$&$ 3$ & $46.8,106.7$&\\
$45$&$25.39$&$-0.20$&$ 14.5$&$ 10.6$&$-64$&$ 25.0\pm 7.2$&$ 43.1\pm 13.0$&$ 3$ & $61.4,109.5$&{\small W42,Ke72}\\ 
$46$&$25.22$&$ 0.31$&$ p $&$ p $&$ 18$&$ 8.0\pm 5.8$&$ 8.0\pm 5.8$&$ 0$ & $44.3,105.5$&{\small RCW173}\\ 
$47$&$24.58$&$ 0.17$&$ 33.9$&$ 10.1$&$ 19$&$ 24.0\pm 7.8$&$ 72.5\pm 24.3$&$ 3$ & $108.0$&{\small W42}\\
$48$&$24.55$&$-0.16$&$ 29.2$&$ 7.1$&$ -2$&$ 25.4\pm 7.9$&$ 62.1\pm 20.2$&$ 3$ & $102.2$&{\small W42}\\ 
$49$&$23.62$&$ 0.22$&$ 17.1$&$ 8.6$&$ 53$&$ 19.1\pm 5.7$&$ 33.6\pm 10.5$&$ 3$ & $90.9$&\\ 
$50$&$23.43$&$-0.17$&$ 30.6$&$ 5.2$&$-30$&$ 22.4\pm 5.8$&$ 54.8\pm 15.0$&$ 3$ & $60.2,93.0$&{\small W41}\\ 
$51$&$23.03$&$ 0.60$&$ p $&$ p $&$ 0$&$ 2.0\pm 2.2$&$ 2.0\pm 2.2$&$ 0$ & $27.7,96.6$&{\small S57}\\ 
$52$&$22.93$&$-0.35$&$ 24.7$&$ 15.9$&$-13$&$ 23.3\pm 5.3$&$ 66.7\pm 16.3$&$ 3$ & $75.4$&{\small W41}\\
$53$&$22.40$&$-0.00$&$ 12.4$&$ p $&$ 90$&$ 7.6\pm 2.2$&$ 7.6\pm 2.2$&$ 2$ &$ 85.0$ &\\ 
$54$&$22.22$&$-0.17$&$ 7.7 $&$ p $&$ 6$&$ 8.5\pm 2.3$&$ 8.5\pm 2.3$&$ 3$ &$ 82.1$ &\\ 
$55$&$21.78$&$-0.17$&$ 17.9$&$ 14.2$&$ 37$&$ 8.0\pm 1.8$&$ 17.4\pm 4.3$&$ 0$ & $79.5$&{\small RCW168}\\
$56$&$20.74$&$-0.10$&$ 13.6$&$ 8.1$&$-14$&$ 17.1\pm 3.3$&$ 26.5\pm 5.6$&$ 0$ & $52.8$&{\small Ke68}\\ 
$57$&$20.27$&$-0.77$&$ p $&$ p $&$ 90$&$ 2.9\pm 0.7$&$ 2.9\pm 0.7$&$ 0$ &$48.1$&\\ 
\hline
\end{tabular}
\end{table*}

%%%%%%%%%%%%%%%%%%%%%%%%%%%%%%%%%%%%%%%%%%%%%%%%%%%%%%%%%%%%%%%%%%
\subsubsection{Properties of the catalogue}
\label{sec:catprop}

From Section \ref{sec:finaldata} it follows that the RRL integrated fluxes for a given source are accurate to the 10 per cent level. However, the $\sim 500$\,K uncertainty on the $T_{\rm e}$ relationship in equation (\ref{eq:te}) means that the integrated fluxes in Table \ref{tab:sextparam_hii} are accurate to the 15 per cent level. 
As an example, we compare the present result for W40 with the higher resolution \citet{Paladini:2003} catalogue, which lists the flux density at 2.7\,GHz for 264 \hii~regions in this longitude range. In \citeauthor{Paladini:2003} the flux density and size of W40 are $34.0 \pm 3.4$\,Jy and $5.5 \pm 2.0$\,arcmin, respectively. We assume this \hii~region is optically thin at 1.4\,GHz based on the spectral index $\alpha=0.0$, where $S \propto \nu^{\alpha}$, given by the \citet{Altenhoff:1970} flux measurements at 1.4, 2.7 and 5\,GHz with a $\sim 10$\,arcmin beam in each case. Therefore, the flux density at 1.4\,GHz extrapolated using the optically thin spectral index $\alpha = -0.1$ is $37.3 \pm 3.6$\,Jy, compared to $ 30.6\pm 5.4$ from this work. The two results are consistent. 
%The difference is caused by the fact that the electron temperature used in the conversion of RRL line integral into free-free temperature is $7680$\,K, whereas the derived $T_{\rm e}$ for this object listed in Table \ref{table:te} is $8770 \pm 560$\,K. This generates a flux density 1.16 times lower, thus the $T_{\rm e}$-corrected flux is $35.5 \pm 6.3$\,Jy which is consistent with the flux from the \citet{Paladini:2003} catalogue.

In a complex of \hii~regions the 14.8\,arcmin beam cannot resolve the individual components, thus comparing the total integrated flux with the individual flux densities from the \citet{Paladini:2003} is not straightforward. The 2.7\,GHz survey by \citet{Reich:1990a}, at a resolution of 4.3\,arcmin, was used to identify the individual \hii~regions within the source G26.66-0.14 listed in Table \ref{tab:sextparam_hii}. These are G26.6-0.1 and G26.6-0.3, listed in the \citeauthor{Paladini:2003} catalogue with angular sizes of $5.5 \pm 1.9$\,arcmin and $11.0 \pm 3.6$\,arcmin, and velocities $104.2 \pm 1.8$\kms~and $69.2 \pm 1.6$\kms, respectively. 
%Their flux densities extrapolated to 1.4\,GHz using $\alpha=-0.1$ are $2.8 \pm 0.5$\,Jy and $15.0 \pm 5.0$\,Jy. 
The present RRL spectrum at $(\ell,b)=(26\fdg66,-0\fdg14)$ has two velocity components, 101.7 and 73.0\kms,  which means that the emission is indeed arising from these two \hii~regions.
r.

%%%%%%%%%%%%%%%%%%%%%%%%%%%%%%%%%%%%%%%%%%%%%%%%%%%%%%%%%%%%%%%%%%
\subsection{The latitude distribution of individual \hii~regions}
\label{sec:hiilat}

\begin{figure*}
\includegraphics[scale=0.5]{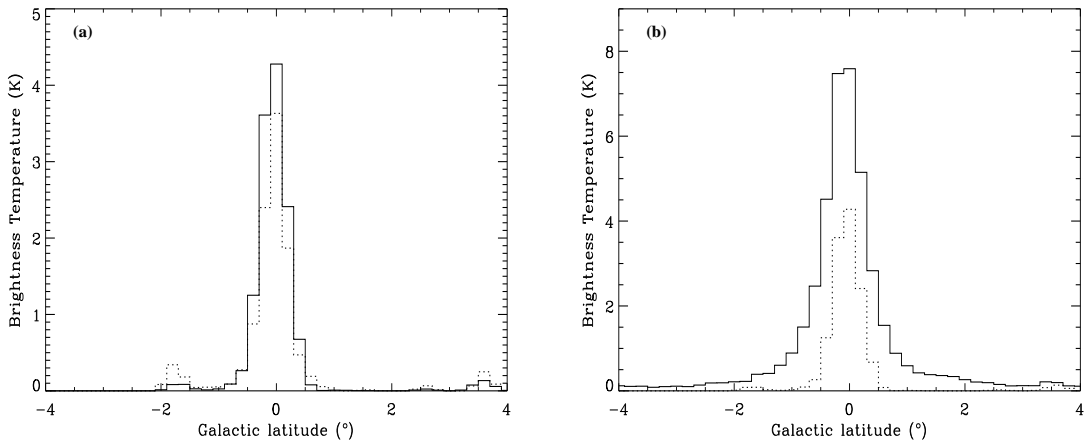}
\caption{(a) Comparison between the contribution from the 159 \citet{Paladini:2003} sources (dotted line) and the 57 sources from extracted from this work (full line), versus latitude. (b) Separation between the diffuse emission and the emission from individual \hii~regions at 1.4\,GHz. The full line represents the total free-free emission and the dotted is the contribution from the SExtractor \hii~regions. Both plots are for the longitude range $20$\degr~to $44$\degr~with $0\fdg2$~latitude bins.}
\label{fig:hiilatcomp}
\end{figure*}

In this Section we compare the latitude distribution of the \hii~regions from the higher resolution \citet{Paladini:2003} catalogue with that from the present, lower resolution catalogue. There are 159 objects in \citeauthor{Paladini:2003} with RRL velocity measurements, in this longitude range. The listed 2.7\,GHz flux densities are extrapolated to 1.4\,GHz assuming the free-free is optically thin therefore well approximated by a power law with $\alpha = -0.1$.
The flux densities and sizes from both catalogues are used to simulate Gaussian profiles and to create the spatial distribution of the \hii~regions in the form of brightness temperature maps at 1.4\,GHz and 14.8\,arcmin resolution. 
We have also estimated the contribution of UC\hii~regions to the total free-free emission in the area under study using the apparent brightness temperature of the 220 catalogued compact \hii~regions at 1.4\,GHz and 5\,arcsec resolution by \citet{Giveon:2005}. About 20 per cent of the objects are likely to be optically thick at this frequency. However the contribution from the UC\hii~regions is found to be relatively low, $1-2$ per cent. Thus the optically thin assumption for the free-free emission is not significantly affected by the presence of these compact objects.
 The UC objects contribute with $1-2$ per cent to the total free-free emission, therefore they is not likely to affect the  With Therefore the contribution of UC objects is small to affect the optically thin assumption on the spectral index of free-free emission. 
%These are smoothed to 14.8\,arcmin resolution and converted into brightness temperature using the conversion factor 0.84\,K/Jy. 
The comparison is shown in Fig. \ref{fig:hiilatcomp} (a) as a function of latitude, averaged over the whole longitude range. The two distributions are very similar, with FWHM of $0\fdg48$ and $0\fdg53$ for the \citeauthor{Paladini:2003} and the present lists, respectively. The catalogue from this work includes the more extended emission around and between features which is not included in the higher resolution \citet{Paladini:2003} catalogue, which is missing $\sim 20\%$ of the flux.

Fig. \ref{fig:hiilatcomp} (b) compares the latitude distribution of the \hii~regions recovered with SExtractor with that of the total free-free emission, in the longitude range $20$\degr~to $44$\degr. It can be seen that the \hii~regions are a narrower distribution than the total diffuse emission, with about half the FWHM. The individual \hii~regions account for $\sim 30\%$ of the total free-free emission in this region of the Galaxy. This result is similar to that found in \paper~for the smaller longitude range $\ell=36$\degr~to 44\degr.

%%%%%%%%%%%%%%%%%%%%%%%%%%%%%%%%%%%%%%%%%%%%%%%%%%%%%%%%%%%%%%%%%%
%%%%%%%%%%%%%%%%%%%%%%%%%%%%%%%%%%%%%%%%%%%%%%%%%%%%%%%%%%%%%%%%%%
\section{DERIVATION OF A 1.4\,GHz SYNCHROTRON TEMPLATE}
\label{sec:synctemp}

The synchrotron emission at 1.4\,GHz for the region $\ell=20$\degr~to 44\degr~and $|b| \le 4$\degr~can now be obtained by subtracting the final free-free emission from the total continuum. The result in show in Fig. \ref{fig:maps}. Well-known SNRs are identified in the map along with a broad background emission falling to $\sim 4$\,K at $|b| \ltsim 4$\degr.
There are some positive and negative residuals at the position of a few \hii~regions in the synchrotron map. W40 and W48 are likely to be underestimated in free-free emission caused by the adoption of an electron temperature for the \hii~region less than the value used for the diffuse emission. In the case of W40, the broad emission underlying the compact source is thought to be of non-thermal origin. The opposite applies to W47 and W43. These are groups of several \hii~regions, unlikely to be characterised by a single $T_{\rm e}$ value.

\begin{figure}
\hspace{-0.5cm}
\includegraphics[scale=0.5]{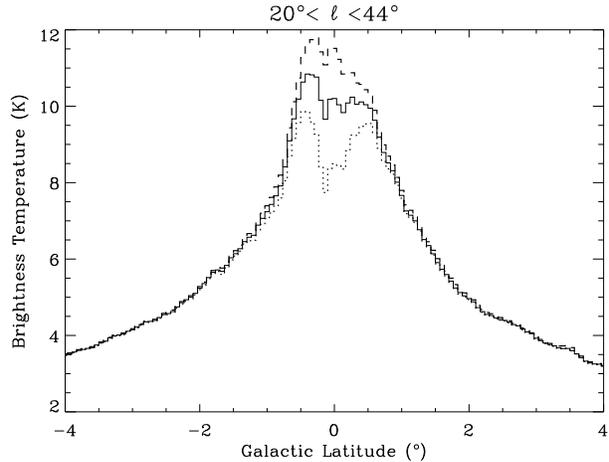}
\caption{The synchrotron emission versus latitude, averaged over the 24\degr~$\ell$-range using 4\,arcmin latitude bands, obtained with the $T_{\rm e}-R_{\rm G}$ relationship (full), $T_{\rm e}=7000$\,K (dotted) and $T_{\rm e}=5000$\,K (dashed).}
\label{fig:syncdist}
\end{figure}

Fig. \ref{fig:syncdist} gives the synchrotron emission versus latitude, averaged over the $24$\degr~range in longitude, using 4\,arcmin latitude bands. The three lines are the different synchrotron results using $T_{\rm e}=T_{\rm e}(R_{\rm G})$, $T_{\rm e}=7000$\,K and $T_{\rm e}=5000$\,K in the free-free estimation. 
The similarity between the three plots for $|b| \gtsim 1$\degr~reflects the fact that the RRL line integral is low away from the plane, and therefore the synchrotron emission is the major contributor at 1.4\,GHz for $|b| \gtsim 1$\degr.
The contribution of the strongest SNR in this region, W44 (G34.7-0.4), is visible at $b \approx -0\fdg4$. 
The smoother distribution given by the full and dashed lines in Fig. \ref{fig:syncdist}, with relation to the dotted line, seems to indicate that $T_{\rm e}$ of the diffuse ionised gas on the Galactic plane is similar to that of the individual \hii~regions and not higher.
The synchrotron distribution appears to have two components - one with a FWHM of $\sim 10$\degr~and a narrow Galactic plane component with a FWHM of $\sim 2$\degr. This narrow component is identified here for the first time after direct subtraction of the free-free emission and is presumably arising from the SNRs on the plane which have different ages.

%%%%%%%%%%%%%%%%%%%%%%%%%%%%%%%%%%%%%%%%%%%%%%%%%%%%%%%%%%%%%%%%%%
\subsection{Catalogue of synchrotron features}
\label{sec:synccat}

The synchrotron map derived in this work, as the best measure of synchrotron emission at 1.4\,GHz, enables the detection of synchrotron features, most of which are SNRs. SExtractor was used to create a list of SNRs in the region $\ell=44$\degr~to $20$\degr. 
The background mesh size, the filter function and the detection minimum area are the same as for the free-free map but the detection threshold is decreased from 2 to $1.5 \sigma$. The rms of the resulting background map is 0.69\,Jy/beam and the threshold limit for a detection is therefore 1.03\,Jy/beam. A Gaussian filter was also used in order to detect diffuse and extended such as the W50 SNR, not easily detected by the Mexican hat filter. Table \ref{tab:sextparam_snr} gives the angular size, position angle, peak and total flux density for the 26 SNRs recovered by SExtractor.

\begin{table*}
\caption{SNRs in the region $\ell=44$\degr~to $20$\degr, extracted from the present synchrotron map using SExtractor. The column descriptions are the same as in Table \ref{tab:sextparam_hii}. The last column identifies a source by its commonly used name as in the \citet{Green:2009} catalogue. The Flag parameter 16 for the first and last SNR listed means that the objects are close to the image boundary. The 7 objects with the star symbol before the number are those extracted using a Gaussian filter instead of the Mexican hat filter.
\label{tab:sextparam_snr}}
\begin{tabular}{cccccccccc}
\\
\hline
\hline
{\small Number} & $\ell$ & $b$  & $\theta_{a}$ & $\theta_{b}$ & {\small P.A.} & {\small Peak} & {\small Flux} & {\small Flag} &{\small Name}\\
& ($^{\circ}$) & ($^{\circ}$) & ($^{\prime}$) & ($^{\prime}$) & ($^{\circ}$) & {\small (Jy/beam)}  & {\small (Jy)} & &   \\
%& \\
\hline
$ ^{\star} 1$&$43.98$&$ 1.60$  &$ 35.4$&$ 25.4$&$-77$&$   2.0\pm   1.1$&$  10.1\pm   5.8$&$  16.0$  &\\
$          2$&$43.23$&$-0.13$ &$ 16.0$&$ p   $&$ 58$&$   29.8\pm   1.3$&$  49.7\pm   5.5$&$   0.0$ & {\small W49B}\\
$ ^{\star} 3$&$42.12$&$-0.21$  &$ 51.34$&$41.1$&$ 78$&$   3.8\pm   1.3$&$  40.3\pm  14.4$&$   1.0$  &\\
$ ^{\star} 4$&$41.45$&$ 0.39$  &$ 26.8$&$ 20.3$&$  1$&$   4.3\pm   1.4$&$  15.0\pm   5.1$&$   1.0$  & \\
$          5$&$41.10$&$-0.30$ &$   p$&$ p   $&$ -3$&$   13.3\pm   1.5$&$  16.0\pm   1.8$&$   0.0$ & {\small 3C397}\\
$ ^{\star} 6$&$40.58$&$-0.45$  &$ 34.0$&$ 28.0$&$-13$&$   5.0\pm   1.6$&$  26.9\pm   9.2$&$   3.0$  & \\
$ ^{\star} 7$&$39.78$&$-2.34$  &$113.4$&$ 53.2$&$ 73$&$   4.9\pm   1.3$&$ 141.3\pm  40.3$&$  17.0$  & {\small W50}\\
$          8$&$39.22$&$-0.32$ &$ 4.2$&$ p    $&$-14$&$   12.7\pm   1.8$&$  12.7\pm   1.8$&$   0.0$ & {\small 3C396}\\
$ ^{\star} 9$&$36.61$&$-0.83$  &$ 21.0$&$ p   $&$-51$&$   2.3\pm   1.4$&$   4.0\pm   2.4$&$   0.0$  & \\
$         10$&$35.59$&$-0.44$ &$ p   $&$ p    $&$ 55$&$   6.7\pm   1.5$&$   6.7\pm   1.5$&$   0.0$ &\\
$         11$&$34.68$&$-0.44$ &$ 26.0$&$ 15.4$&$-56$&$   71.2\pm   1.6$&$ 208.1\pm  18.6$&$   0.0$ & {\small W44}\\
$         12$&$33.67$&$ 0.06$ &$ p   $&$ p   $&$ 43$&$   10.3\pm   1.6$&$  10.3\pm   1.6$&$   0.0$ & {\small Kes79}\\
$ ^{\star}13$&$33.21$&$-0.51$  &$ 7.9 $&$ p   $&$ 14$&$   2.3\pm   1.4$&$   2.3\pm   1.4$&$   0.0$  &\\
$         14$&$32.86$&$-0.05$ &$  p  $&$  p  $&$ -1$&$    6.5\pm   1.5$&$   6.5\pm   1.5$&$   0.0$ & {\small Kes78}\\
$         15$&$31.89$&$ 0.04$ &$  3.9$&$ p   $&$ 25$&$   14.7\pm   1.7$&$  14.7\pm   1.7$&$   0.0$ & {\small 3C391}\\
$         16$&$29.69$&$-0.27$ &$ p   $&$ p   $&$ 80$&$    7.9\pm   2.0$&$   9.5\pm   2.4$&$   0.0$ & {\small Kes75}\\
$         17$&$28.61$&$-0.05$ &$ p   $&$  p  $&$-72$&$    8.6\pm   1.8$&$   8.6\pm   1.8$&$   0.0$ &\\
$         18$&$27.75$&$ 0.60$ &$ p   $&$  p  $&$  0$&$   10.9\pm   1.4$&$  10.9\pm   1.4$&$   0.0$ &\\
$         19$&$27.37$&$ 0.00$ &$ p   $&$ p  $&$-51$&$    8.9\pm   1.3$&$  10.7\pm   1.6$&$   3.0$ & {\small 4C-04.71}\\
$         20$&$24.73$&$-0.70$ &$  p  $&$  p  $&$ 90$&$    5.4\pm   1.5$&$   5.4\pm   1.5$&$   0.0$ &\\
$         21$&$24.58$&$ 0.62$ &$ 10.1$&$ p   $&$ 23$&$   11.4\pm   1.7$&$  12.7\pm   3.2$&$   0.0$ &\\
$         22$&$23.17$&$-0.24$ &$ 8.8 $&$ p   $&$ 81$&$   13.0\pm   2.3$&$  14.5\pm   4.1$&$   2.0$ & {\small W41}\\
$         23$&$22.67$&$-0.24$ &$ 17.3$&$ 2.4 $&$ 73$&$   15.5\pm   2.0$&$  24.1\pm   5.1$&$   0.0$ &\\
$         24$&$21.81$&$-0.57$ &$ 14.5$&$ 7.3 $&$ 34$&$   32.5\pm   2.0$&$  51.0\pm   6.2$&$   0.0$ & {\small Kes69}\\
$         25$&$21.50$&$-0.87$ &$  p  $&$  p  $&$  0$&$    8.2\pm   2.1$&$   9.8\pm   2.5$&$   0.0$ &\\
$         26$&$19.94$&$-0.21$ &$ 17.1$&$ p   $&$-47$&$   12.0\pm   1.7$&$  16.4\pm   4.3$&$  16.0$ &\\
\hline
\end{tabular}
\end{table*}

The latest published compilation of Galactic SNRs is given by \citet{Green:2009}. It contains 36 SNRs in the $\ell$-range under study, 23 of which are detected in this survey and listed in Table \ref{tab:sextparam_snr}. Objects 3, 4 and 10, are not in the \citeauthor{Green:2009} catalogue and are described in more detail in the next section.
Most of the 13 SNRs in the \citeauthor{Green:2009} catalogue not detected in this work have flux densities below 2.7\,Jy, which is comparable with the lowest flux density in Table \ref{tab:sextparam_snr}. Moreover some of the SNRs are poorly defined, without clear angular sizes or flux measurements. 

Most of the sources in Table \ref{tab:sextparam_snr} are smaller than the beam, their sizes and flux densities are in overall agreement with the values in the \citeauthor{Green:2009} catalogue. For example the strongest SNR in this region of the Galaxy, W44, is fitted in the present synchrotron map with a size of $26\farc.0 \times 15\farc.4$ and a flux density of $208 \pm 19$\,Jy. The flux density given by \citet{Green:2009} for this object is 203\,Jy at 1.4\,GHz, extrapolated from 1\,GHz with the listed spectral index $\alpha=0.37$, and the angular size is $35^{\prime} \times 27^{\prime}$.

%%%%%%%%%%%%%%%%%%%%%%%%%%%%%%%%%%%%%%%%%%%%%%%%%%%%%%%%%%%%%%%%%%
\subsection{Comments on individual synchrotron sources}
\label{sec:commsync}

%%%%%%%%%%%%%%%%%%%%%%%%%%%%%%%%%%%%%%%%%%%%%%%%%%%%%%%%%%%%%%%%%%
\subsubsection{G42.12-0.21 and G41.45+0.39}
\label{sec:snrsources1}
These objects correspond to two of the three SNR candidates by \citet{Kaplan:2002}. The angular sizes and flux densities found here are larger than the values obtained by \citet{Kaplan:2002} using the 1.4\,GHz NRAO VLA Sky Survey (NVSS, \citealt{Condon:1998}) data, which may resolve out some of the emission. This can also be due to the fact that these two faint objects are recovered with the Gaussian filter, thus are likely to be confused with the strong background on the plane. Moreover, they are both flagged as having bright sources nearby. The third possible SNR in \citet{Kaplan:2002}, G43.5+0.6, is too faint in the NVSS to derive a flux density and it is indeed only marginally visible, at a $\sim 1.2 \sigma$ level, in the present synchrotron map. 

%%%%%%%%%%%%%%%%%%%%%%%%%%%%%%%%%%%%%%%%%%%%%%%%%%%%%%%%%%%%%%%%%%
\subsubsection{G35.6-0.4}
\label{sec:snrsources2}
This SNR has been recently re-identified by \citet{Green:2009b} using VLA Galactic Plane Survey (VGPS, \citealt{Stil:2006}) data and data from other single-dish surveys at higher frequency to confirm the non-thermal spectral index of this object. G35.6-0.4 is seen in the VGPS map with a size of $\sim 15^{\prime} \times 11^{\prime}$ and has an flux density of 7.8\,Jy. These results are consistent with what is found in this work, $6.7 \pm 1.5$\,Jy for a relatively compact source of size less than or equal to that of the beam.
\\

We have thus confirmed the synchrotron nature of the objects G42.12-0.21, G41.45+0.39 and G35.6-0.4 using the present survey.

The similar resolution of the 2.7\,GHz survey by \citet{Reich:1990a} and the 100\,$\mu$m data by \citet{Miville-Deschenes:2005} allows the investigation of the spatial coincidence between dust and radio continuum emission, used to distinguish between \hii~regions and SNRs \citep{Haslam:1987,Broadbent:1989}. This correspondence helps us understand that the artifacts seen in the synchrotron map around the \hii~region complexes W47, W43 and W42 are correlated with individual \hii~regions, and are thus caused by electron temperature variations. On the other hand, the lack of significant dust emission for the sources at $(\ell,b)=(30\fdg1,+1\fdg3)$ and $(\ell,b)=(21\fdg0,+2\fdg0)$ in the synchrotron map, points towards a non-thermal origin. Both sources are extragalactic, even though only $(\ell,b)=(21\fdg0,+2\fdg0)$ is currently identified as such in the literature \citep{Clark:1974}. 

\begin{figure*}
\includegraphics[scale=0.55]{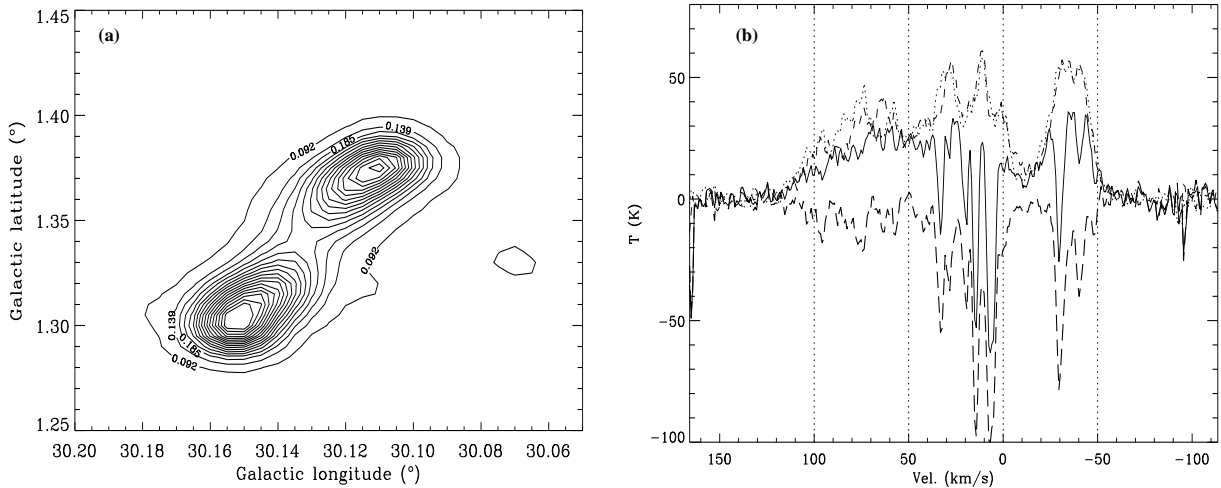}
\caption{(a) Contour map of the radio source $(\ell,b)=(30\fdg13,+1\fdg34)$ from the VGPS survey at 1.4\,GHz and 1\,arcmin resolution. The contours are given at every $5 \%$ from 5 to $90\%$ of 0.92\,Jy/beam, which corresponds to the maximum on the southern lobe. The first 4 contours are labelled. (b) The \hi~spectrum from the VGPS survey against the radio source, taken at $(\ell,b)=(30\fdg15,1\fdg31)$ (full line). The dotted and dashed lines are the off source spectra, measured at the same $b$ and $\ell=31\fdg13$ and $31\fdg17$, respectively. The thick long-dashed line is the absorption spectrum of the source which shows that absorption is detected at all velocities, inside and outside the Solar circle, confirming its extragalactic origin.}
\label{fig:gal}
\end{figure*}

%%%%%%%%%%%%%%%%%%%%%%%%%%%%%%%%%%%%%%%%%%%%%%%%%%%%%%%%%%%%%%%%%%
\subsubsection{The double-lobed radio source $(\ell,b)=(30\fdg13,+1\fdg34)$}
\label{sec:radiogal}

The source at $(\ell,b)=(30\fdg13,+1\fdg34)$ has been detected in previous surveys \citep{Reich:1990a,Reich:1990,Slee:1995,Cohen:2007} but not identified. We use VGPS \hi~data to identify this object a radio galaxy.
The contour map of Fig. \ref{fig:gal} (a) is the VGPS 1.4\,GHz emission of the radio source. At this survey of resolution 1\,arcmin a double-peaked feature is clearly visible. The elongated structure has been detected at lower frequencies, in particular at 74\,MHz in the VLA Low-Frequency Sky Survey (VLSS, \citealt{Cohen:2007}). The VLSS catalogue gives a flux density of $41.97 \pm 4.24$\,Jy and an angular size of $2\farc.05 \times 0\farc.59$ with a 1.3\,arcmin beam. At 1.4\,GHz, the survey by \citet{Reich:1990} at 9.4\,arcmin resolution gives a flux density of $4.62 \pm 0.05$\,Jy for a slightly extended source of angular size less than 11\,arcmin. The higher resolution survey at 2.7\,GHz \citep{Reich:1990a} detects a source of $7\farc.4 \times 4\farc.3$ size and flux density of $3.57 \pm 0.04$\,Jy. Combining the three measurements results in a spectral index of $\alpha=-0.71 \pm 0.05$, which is within the observed values for a SNR \citep{Green:2009}. Nevertheless, the shape of the radio source and its angular size points toward an extragalactic origin. 
\citet{Helfand:1992}, on their 20\,cm study with the VLA and identification of 1457 compact sources near the Galactic plane, also suggest that this is an extragalactic object. If this is the case, such a bright and extended object, has probably been missed from extragalactic source catalogues due to its proximity to the Galactic plane. 

The latitude coverage of the VGPS \hi~spectral cube, $|b| \ltsim 1\fdg32$, only includes the southern lobe of the source, which is enough to obtain and compare spectra on and off the source. These are shown in Fig. \ref{fig:gal} (b). \hi~absorption against the object is detected at all velocities including negative velocities, which gives the confirmation that it is outside the Galaxy.

This interesting object deserves further study at higher angular resolution.

%%%%%%%%%%%%%%%%%%%%%%%%%%%%%%%%%%%%%%%%%%%%%%%%%%%%%%%%%%%%%%%%%%
%%%%%%%%%%%%%%%%%%%%%%%%%%%%%%%%%%%%%%%%%%%%%%%%%%%%%%%%%%%%%%%%%%
\section{COMPARISON WITH THE \textit{WMAP} MAXIMUM ENTROPY MODEL}
\label{sec:mem}

%%%%%%%%%%%%%%%%%%%%%%%%%%%%%%%%%%%%%%%%%%%%%%%%%%%%%%%%%%%%%%%%%%
%\subsection{\textit{WMAP} free-free MEM model}
%\label{sec:mem}

In this Section we extend the analysis performed in \paper~using the three times larger longitude coverage of the present RRL map to compare with the \textit{WMAP} prediction of the free-free emission at 23\,GHz\footnote{Downloaded from the website http://lambda.gsfc.nasa.gov.}.
The Maximum Entropy Model (MEM, \citealt{Gold:2011}) is a spatial and spectral fit that uses external templates in regions of low signal-to-noise. The spectral indices for the free-free and thermal dust emission are fixed, with $\beta=-2.14$ and $\beta=+2.0$, respectively ($T_{b} \propto \nu^{\beta}$). Any additional component of emission such as the anomalous microwave emission is included in the synchrotron component.
To have a reliable prediction of the free-free emission on the Galactic plane is important not only for foregrounds and component separation but also for Galactic studies such as 3-D Galaxy \citep{Planck:2011a} and magnetic field studies \citep{Jaffe:2011}.

\begin{figure}
\hspace{-0.5cm}
\includegraphics[scale=0.5]{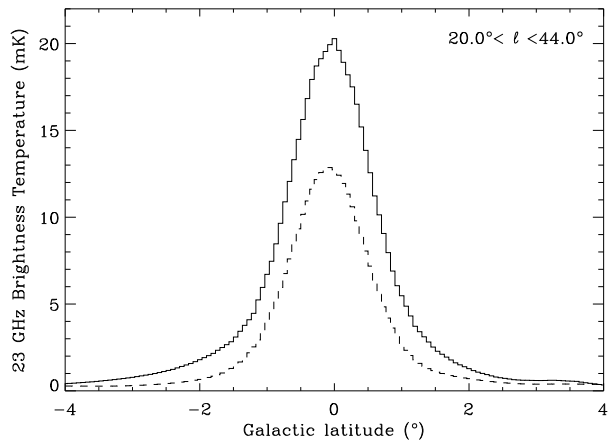}
\caption{Free-free emission estimated from the RRLs (dashed line) and from the \textit{WMAP} 23\,GHz MEM (full line) versus latitude, averaged over the 24\degr~longitude range.}
\label{fig:mem}
\end{figure}

In \paper~the comparison between the free-free estimated from the RRLs and from the \textit{WMAP} seven year MEM suggested that the electron temperature of the ionised medium in the $\ell$-range 36\degr~to 39\degr~is about 8000\,K. This is based on the higher MEM distribution on the Galactic plane, by 30 per cent, with relation to the prediction from the RRLs.
Fig. \ref{fig:mem} shows the comparison between the two datasets using the extended longitude range $\ell=20$\degr~to 44\degr~from the present survey, at 23\,GHz. The spectral index used to extrapolate the free-free brightness temperature from the RRLs is $\beta=-2.13$ which is the average value across the range of electron temperatures given by equation (\ref{eq:te}). The free-free spectral index $\beta$ is a slow function of frequency and electron temperature \citep{D3:2003}, decreasing as the frequency increases. The steeper spectral index used in the MEM analysis reflects this behaviour. The RRL data are smoothed to the MEM resolution of 1\degr. The FWHM of the MEM latitude distribution is $1\fdg48$, slightly broader then that of the RRLs, $1\fdg36$, with a $\sim 50$~per cent higher peak. 
The free-free brightness temperature depends on $T_{\rm e}^{1.15}$ (Section \ref{sec:introduction}). Therefore this difference implies an electron temperature of $\sim 10000$\,K for the ionised gas on the Galactic plane, compared with the mean 6080\,K used in the RRL conversion to free-free (Section \ref{sec:ff}). Based on the results presented in previous sections, we expect the electron temperature of the diffuse to be around 6000\,K in this region of the Galactic plane.

Similar results are reported by \citet{Jaffe:2011}. They recover 80 per cent or less of the MEM free-free emission in their model of the Galactic magnetic field.
Results from the \textit{Planck} satellite also indicate that the MEM overpredicts the ionised component on the Galactic plane \citep{Planck:2011a}. In this case, the MEM estimation is $\sim 20$~per cent higher than that recovered from the Galactic inversion model performed. 

The fact that the MEM model does not fit for a separate anomalous microwave emission (AME) template may contribute to the excess deduced for the free-free template relative to the RRL observations. Furthermore, we note that the use of a flatter spectral index for the dust emission, with respect to $\beta=+2.0$, would allow a lower free-free contribution at lower frequencies.

The ratio between the MEM and RRL predictions of the free-free emission varies with longitude. The relative contribution of these two diagnostics will be investigated in a future paper, using the full ZOA Galactic plane coverage.

%%%%%%%%%%%%%%%%%%%%%%%%%%%%%%%%%%%%%%%%%%%%%%%%%%%%%%%%%%%%%%%%%%
%%%%%%%%%%%%%%%%%%%%%%%%%%%%%%%%%%%%%%%%%%%%%%%%%%%%%%%%%%%%%%%%%%
\section{THE 3-D GALAXY}
\label{sec:3dgal}

The present RRL survey is unique for it provides information on the distribution of individual \hii~regions and the diffuse ionised gas on the Galactic plane.

%%%%%%%%%%%%%%%%%%%%%%%%%%%%%%%%%%%%%%%%%%%%%%%%%%%%%%%%%%%%%%%%%%
\subsection{The $\ell-V$ diagram}
\label{sec:lv}

\begin{figure}
\center
\includegraphics[scale=0.7]{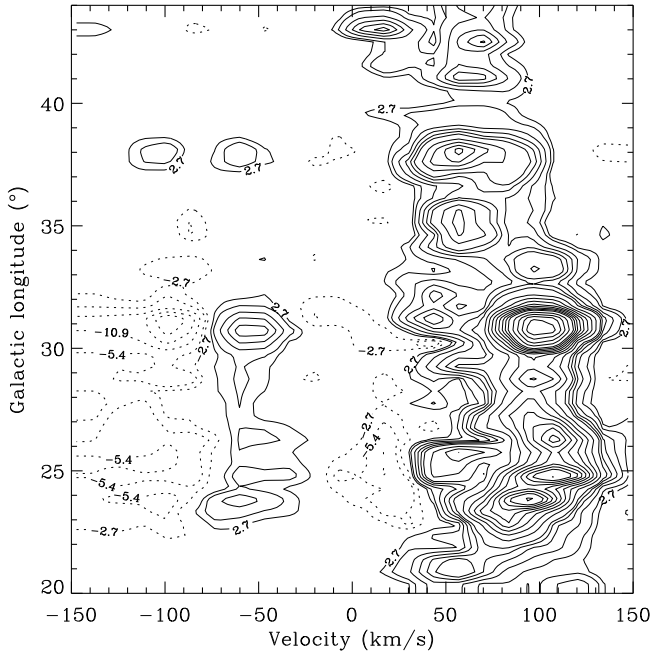}
\caption{The longitude-velocity distribution of RRL brightness temperature within 1\degr~of the Galactic plane. Contours are at $-8$, $-6$, $-4$, $-2$, $-1$, 1, 2, 4, 6, 8 and 10 per cent, then at every 5 per cent from 10 to 50 per cent and every 10 per cent from 60 to 90 per cent of the maximum line temperature observed, 272\,mK. The first 6 contours are labelled and the dotted lines correspond to the negative contours.}
\label{fig:thislonvel}
\end{figure}

Fig. \ref{fig:thislonvel} shows the contours of RRL temperature in longitude-velocity space, for spectra averaged in $0\fdg5$ in longitude within $1$\degr~of the Galactic plane. The first positive contour (full line) is at $\sim 1\sigma$. It shows that there is no H RRL emission at negative velocities in this longitude range: the two vertical bands of emission at $V < 0$\kms~are associated with the He/C RRLs (Section \ref{sec:finaldata}). 
This result is very similar to that found in the H166$\alpha$~line survey by \citet{Lockman:1976} at a resolution of 21\,arcmin, which detected emission in all the $1$\degr~intervals from $4$\degr~to $44$\degr. Both $\ell-V$ maps show a minimum of emission at $\ell \sim 40$\degr, no emission at 0\kms~and a large change in emission with longitude. 
The main band of emission from the H RRLs shows evidence of a slope in longitude, also seen in the \hi~$\ell-V$ diagram in \citealt{Lockman:1976}. This is the result of circular rotation around the Galactic centre, by which gas at higher longitude is further away from the centre and thus has lower radial velocity.

%%%%%%%%%%%%%%%%%%%%%%%%%%%%%%%%%%%%%%%%%%%%%%%%%%%%%%%%%%%%%%%%%%
\subsection{The radial distribution of the ionised gas}
\label{sec:raddist}

\begin{figure}
\hspace{-0.5cm}
\includegraphics[scale=0.6]{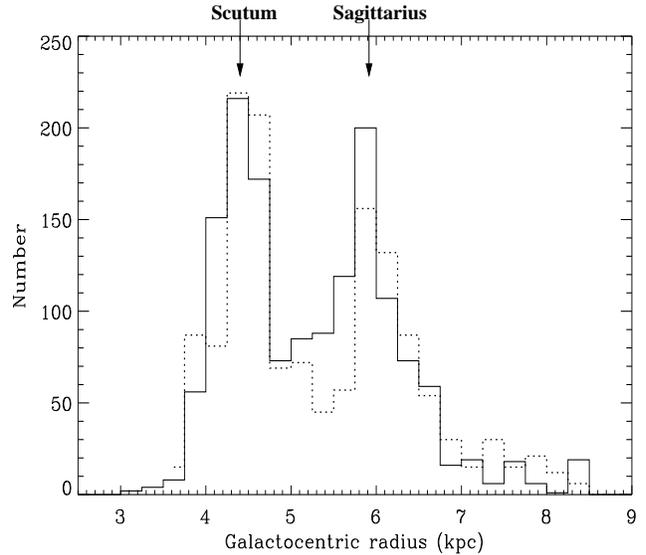}
\caption{Radial distribution of the ionised emission (full line) in the region $\ell=20$\degr~to 44\degr~and $|b| \le 1$\degr. The Galactic distance is calculated for the 1503 discrete RRL components using $R_{0}=8.5$\,kpc. The dotted line is the radial distribution of the 471 \hii~regions from the combination of the \citet{Paladini:2003} and \citet{Anderson:2011} lists.}
\label{fig:raddist}
\end{figure}

The RRL velocity, when converted into a distance to the Galactic centre can be used to obtain the radial distribution of the ionised gas. This is shown in Fig. \ref{fig:raddist}. The RRL spectra are averaged into a 12\,arcmin pixel grid, to reduce the correlation between the pixels, for the longitude range $\ell=20$\degr~to $44$\degr~and $|b| \le 1$\degr. The Galactocentric radius is calculated as in Section \ref{sec:te} using the \citet{Fich:1989} rotation curve, with $R_{0}=8.5$\,kpc. The central velocity of the RRLs is found by fitting a one or two component Gaussian function to the spectra, in the case of a single or double velocity component, respectively. From 1331 lines of sight we detect 1503 discrete RRL components. Fig. \ref{fig:raddist} also shows the distribution of the 471 \hii~regions in this longitude range, from the combination of the \citet{Paladini:2003} and \citet{Anderson:2011} samples (scaled by a factor of 3). The Green Bank Telescope \hii~Region Discovery Survey (HRDS, \citealt{Bania:2010}) has detected 313 new \hii~regions between $\ell=20$\degr~and $44$\degr, adding to the 145 objects with RRL measurements in present region from the \citeauthor{Paladini:2003} list.

\begin{figure*}
\center
\includegraphics[scale=0.45]{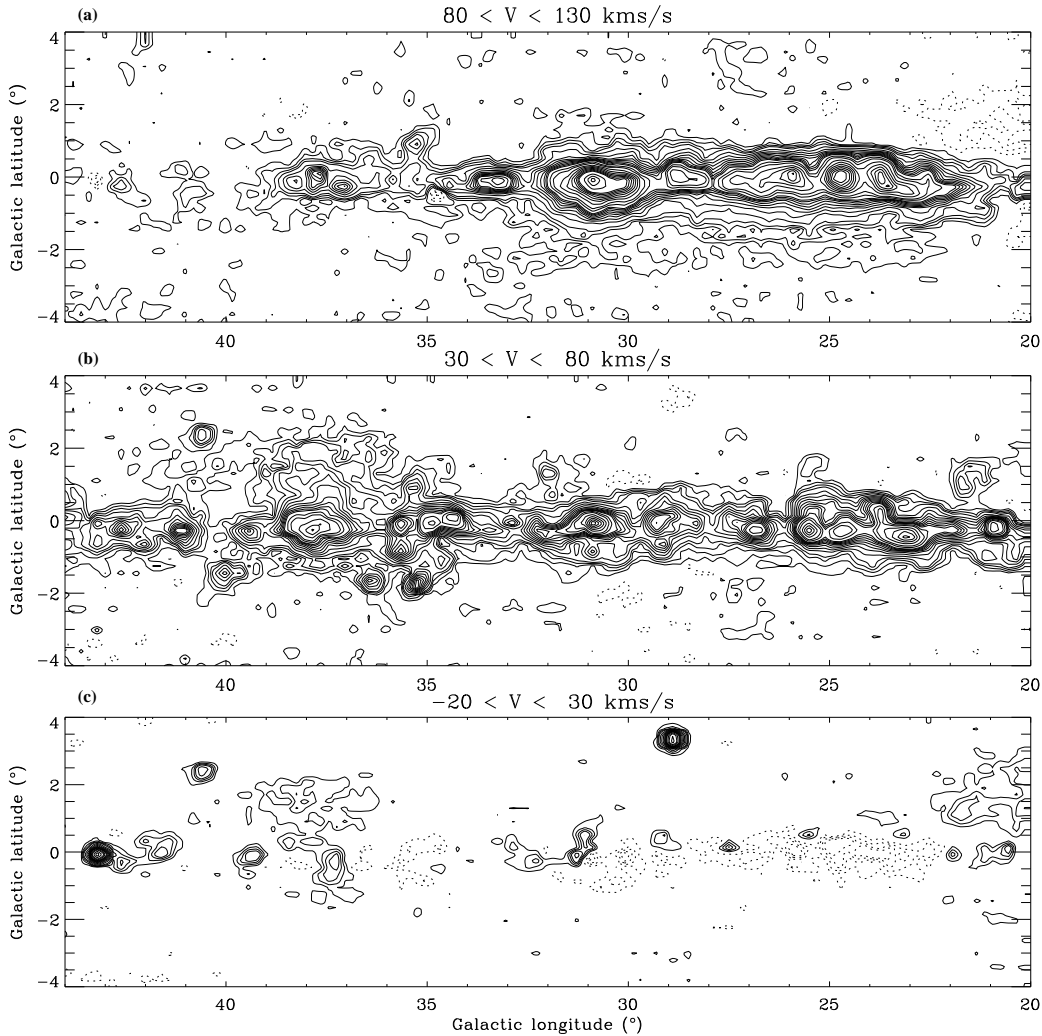}
\caption{Maps of RRL emission integrated over 50\kms, centred at 105\kms~(a), 55\kms~(b) and 5\kms~(c), at a resolution of 14.8\,arcmin. Contours are given at every 0.1\,K\kms~from $-0.4$ to 0.4\,K\kms, every 0.2\,K\kms~from 0.4 to 2\,K\kms~and then at 2.5, 3, 4, 5, 6, 7, 8, 9, 10, 15, 20, 25\,K\kms. The negative contours are dotted.}
\label{fig:chanmaps}
\end{figure*}

The two distributions are very similar and show narrow peaks, $\approx 0.75$\,kpc wide, at $\sim 4.4$\,kpc and $5.9$\,kpc that correspond to the Scutum and Sagittarius spiral arms, respectively.
This confirms the spatial correlation between the diffuse gas and the individual \hii~regions, observed by \citet{Hart:1976} and \citet{Lockman:1976} in the H166$\alpha$ line using pointed observations along the Galactic plane. The $R_{\rm G}$ values of the peaks correspond to tangent points in longitude of $30$\degr~and $45$\degr~and terminal velocities of 110\kms~and 63\kms, respectively. The fact that the present survey covers $\ell \le 44$\degr~suggests that it is missing part of the emission from the tangent point of the Sagittarius arm at $\ell=50$\degr.

%%%%%%%%%%%%%%%%%%%%%%%%%%%%%%%%%%%%%%%%%%%%%%%%%%%%%%%%%%%%%%%%%%
\subsection{The z-distribution of the ionised gas}
\label{sec:zdist}

\begin{figure*}
\center
\includegraphics[scale=0.65]{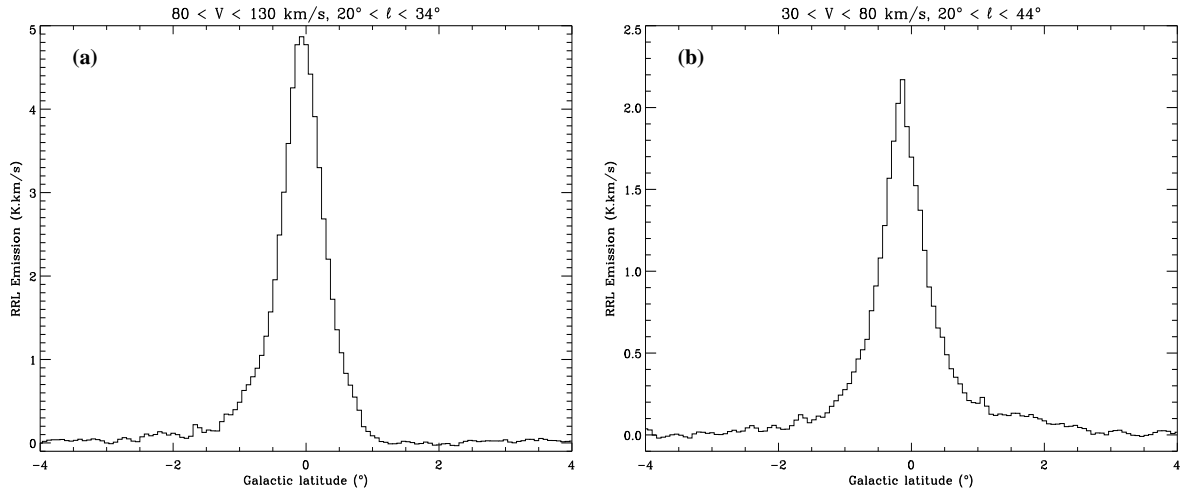}
\caption{The latitude profiles that correspond to maps from Figs. \ref{fig:chanmaps} (a) and (b), respectively, averaged over the longitude ranges indicated.}
\label{fig:chanmaps1}
\end{figure*}

Having identified the Sagittarius and Scutum arms in the $R_{\rm G}$ and velocity space, we are now able to separate the emission from each of the spiral arms and investigate their azimuthal extent. Figs. \ref{fig:chanmaps} (a) and (b) show the integrated RRL emission over 50\kms, from 80 to 130\kms~and from 30 to 80\kms, respectively. Fig. \ref{fig:chanmaps} (a) shows the narrow distribution of the high-velocity emission from the Scutum arm that covers mainly $\ell \ltsim 34$\degr.  The lower emission for $ \ell > 34$\degr~originates in the wings of RRLs at $V \sim 80$\kms. The emission from the Sagittarius arm covers the whole longitude range and has a broader distribution around the plane. Fig. \ref{fig:chanmaps} (c) shows that the RRL emission at $V \sim 0$\kms~is attributed to the W49 and W40 \hii~regions, with some emission from W45, at $V \sim 25$\kms~and from the positive latitude extent of the W47 complex. 

The latitude profiles of Figs. \ref{fig:chanmaps1} (a) and (b) show the narrower z-distribution of the Scutum arm, with a FWHM of $\simeq 0\fdg78$, compared with the slightly broader distribution of the Sagittarius arm, with FWHM of $\simeq 0\fdg85$. The Galactocentric distances to the tangent points give a mean distance to the Sun of 6.0 and 7.3\,kpc for the Sagittarius and Scutum arms, respectively. Using the observed FWHM of the latitude profiles gives a z-thickness of about 100\,pc, $\sigma \sim 45$\,pc, for the diffuse ionised gas about the Galactic plane\footnote{Where $\sigma=\mbox{FWHM}/2\sqrt{2\ln(2)}$}. This compares with $\sigma \sim 70$\,pc, the value found by \citet{Reynolds:1991} using pulsar Dispersion Measures and reflects the fact that emission measures trace the densest parts of the ionised gas, ${\rm EM} \propto n_{\rm e}^{2}$, compared to dispersion measures which trace $n_{\rm e}$.
The latitude distribution of the Sagittarius arm, Fig. \ref{fig:chanmaps1} (b), shows a broader component of FWHM $\simeq 2$\degr~that corresponds to the emission from the near side, whereas emission from the far side is responsible for the narrow component. 
%The negative mean value of the peak in both latitude distributions is consistent with the result from \citet{Reed:1997} that the Sun lies above the Galactic mid-plane.

%%%%%%%%%%%%%%%%%%%%%%%%%%%%%%%%%%%%%%%%%%%%%%%%%%%%%%%%%%%%%%%%%%
%%%%%%%%%%%%%%%%%%%%%%%%%%%%%%%%%%%%%%%%%%%%%%%%%%%%%%%%%%%%%%%%%%
\section{CONCLUSIONS}
\label{sec:conc}

This work has confirmed that the HIPASS/ZOA survey can provide a reliable estimate of the RRL emission on the Galactic plane at 1.4\,GHz and 14.8\,arcmin resolution. The use of the total continuum emission from the same survey has enabled the electron temperature of individual \hii~regions in the area under study to be estimated, showing the well-known Galactocentric gradient consistent with previous results. 

The unambiguous determination of the free-free emission from the present work is used to derive the synchrotron map at 1.4\,GHz. The latitude distribution of the synchrotron emission shows a narrow component on the Galactic plane which is found here for the first time clear of the free-free emission by direct subtraction.

The free-free and synchrotron maps are used to create a list of 57 \hii~regions and 26 SNRs, respectively. The fact that we detect a recently re-identified SNR, G35.6-0.4 \citep{Green:2009b} illustrates the potential of this survey to detect and identify new objects. The contribution of the individual \hii~regions to the total free-free emission is found to be around 30 per cent. Their latitude distribution has a FWHM of $\sim 0\fdg5$, about half the width of the total free-free distribution. This can be converted into a z-thickness for the \hii~emission at known distances and is consistent with previous results on the distribution of the OB stars and the width of the thin disk from Dispersion Measures.

The present estimate, based on the integrated RRLs converted to brightness temperature of the free-free emission recovers about 50 per cent of the \textit{WMAP} MEM prediction. Possible reasons for this discrepancy are likely associated with the thermal dust spectral index and with the fact that there is no separate AME template from MEM.

This study will soon be extended to the inner Galaxy region, $\ell=-50$\degr~to 50\degr, using the improved data analysis pipeline described in Section \ref{sec:datared}. We will then focus on the different applications of this result, namely on the investigation of the spectral energy distribution of the different emission components on the Galactic plane and especially on the AME, as a function of longitude.
Also, the combination of \ha~and RRL data to complement the free-free templates on the Galactic plane, which is the ultimate goal of this project, will be subject of a more detailed study.

%%%%%%%%%%%%%%%%%%%%%%%%%%%%%%%%%%%%%%%%%%%%%%%%%%%%%%%%%%%%%%%%%%
%%%%%%%%%%%%%%%%%%%%%%%%%%%%%%%%%%%%%%%%%%%%%%%%%%%%%%%%%%%%%%%%%%
\section*{ACKNOWLEDGEMENTS}

MIRA is funded by the Funda\c{c}\~{a}o para a Ci\^{e}ncia e Tecnologia (Portugal). CD acknowledges a STFC Advanced Fellowship and an ERC IRG grant under the FP7. The Parkes telescope is part of the Australia Telescope which is funded by the Commonwealth of Australia for operation as a National Facility managed by CSIRO.

%%%%%%%%%%%%%%%%%%%%%%%%%%%%%%%%%%%%%%%%%%%%%%%%%%%%%%%%%%%%%%%%%%
%%%%%%%%%%%%%%%%%%%%%%%%%%%%%%%%%%%%%%%%%%%%%%%%%%%%%%%%%%%%%%%%%%

%\bibliography{cmb_refs}
\bibliographystyle{mn2e}
%\begin{thebibliography}{}

%\end{thebibliography}

%\bibliographystyle{hapj}
\bibliography{rrl}

\begin{thebibliography}{43}
\expandafter\ifx\csname natexlab\endcsname\relax\def\natexlab#1{#1}\fi

\bibitem[{{Altenhoff} {et~al}\mbox{.}(1970){Altenhoff}, {Downes}, {Goad},
  {Maxwell}, \& {Rinehart}}]{Altenhoff:1970}
{Altenhoff} W.~J., {Downes} D., {Goad} L., {Maxwell} A., {Rinehart} R., 1970,
  \aaps, 1, 319

\bibitem[{{Alves} {et~al}\mbox{.}(2010){Alves}, {Davies}, {Dickinson}, {Davis},
  {Auld}, {Calabretta}, \& {Staveley-Smith}}]{Alves:2010}
{Alves} M.~I.~R., {Davies} R.~D., {Dickinson} C., {Davis} R.~J., {Auld} R.~R.,
  {Calabretta} M., {Staveley-Smith} L., 2010, \mnras, 405, 1654

\bibitem[{{Anderson} {et~al}\mbox{.}(2011){Anderson}, {Bania}, {Balser}, \&
  {Rood}}]{Anderson:2011}
{Anderson} L.~D., {Bania} T.~M., {Balser} D.~S., {Rood} R.~T., 2011, ArXiv
  e-prints

\bibitem[{{Bania} {et~al}\mbox{.}(2010){Bania}, {Anderson}, {Balser}, \&
  {Rood}}]{Bania:2010}
{Bania} T.~M., {Anderson} L.~D., {Balser} D.~S., {Rood} R.~T., 2010, \apjl,
  718, L106

\bibitem[{{Barnes} {et~al}\mbox{.}(2001){Barnes}, {Staveley-Smith}, {de Blok},
  {Oosterloo}, {Stewart}, {Wright}, {Banks}, {Bhathal}, {Boyce}, {Calabretta},
  {Disney}, {Drinkwater}, {Ekers}, {Freeman}, {Gibson}, {Green}, {Haynes}, {te
  Lintel Hekkert}, {Henning}, {Jerjen}, {Juraszek}, {Kesteven}, {Kilborn},
  {Knezek}, {Koribalski}, {Kraan-Korteweg}, {Malin}, {Marquarding}, {Minchin},
  {Mould}, {Price}, {Putman}, {Ryder}, {Sadler}, {Schr{\"o}der}, {Stootman},
  {Webster}, {Wilson}, \& {Ye}}]{Barnes:2001}
{Barnes} D.~G. {et~al.}, 2001, \mnras, 322, 486

\bibitem[{{Bertin} \& {Arnouts}(1996)}]{Bertin:1996}
{Bertin} E., {Arnouts} S., 1996, \aaps, 117, 393

\bibitem[{{Broadbent}, {Osborne} \& {Haslam}(1989){Broadbent}, {Osborne}, \&
  {Haslam}}]{Broadbent:1989}
{Broadbent} A., {Osborne} J.~L., {Haslam} C.~G.~T., 1989, \mnras, 237, 381

\bibitem[{{Bronfman} {et~al}\mbox{.}(2000){Bronfman}, {Casassus}, {May}, \&
  {Nyman}}]{Bronfman:2000}
{Bronfman} L., {Casassus} S., {May} J., {Nyman} L.-{\AA}., 2000, \aap, 358, 521

\bibitem[{{Clark} \& {Crawford}(1974)}]{Clark:1974}
{Clark} D.~H., {Crawford} D.~F., 1974, Australian Journal of Physics, 27, 713

\bibitem[{{Cohen} {et~al}\mbox{.}(2007){Cohen}, {Lane}, {Cotton}, {Kassim},
  {Lazio}, {Perley}, {Condon}, \& {Erickson}}]{Cohen:2007}
{Cohen} A.~S., {Lane} W.~M., {Cotton} W.~D., {Kassim} N.~E., {Lazio} T.~J.~W.,
  {Perley} R.~A., {Condon} J.~J., {Erickson} W.~C., 2007, \aj, 134, 1245

\bibitem[{{Condon} {et~al}\mbox{.}(1998){Condon}, {Cotton}, {Greisen}, {Yin},
  {Perley}, {Taylor}, \& {Broderick}}]{Condon:1998}
{Condon} J.~J., {Cotton} W.~D., {Greisen} E.~W., {Yin} Q.~F., {Perley} R.~A.,
  {Taylor} G.~B., {Broderick} J.~J., 1998, \aj, 115, 1693

\bibitem[{{Dickinson}, {Davies} \& {Davis}(2003){Dickinson}, {Davies}, \&
  {Davis}}]{D3:2003}
{Dickinson} C., {Davies} R.~D., {Davis} R.~J., 2003, \mnras, 341, 369

\bibitem[{{Fich}, {Blitz} \& {Stark}(1989){Fich}, {Blitz}, \&
  {Stark}}]{Fich:1989}
{Fich} M., {Blitz} L., {Stark} A.~A., 1989, \apj, 342, 272

\bibitem[{{Giveon} {et~al}\mbox{.}(2005){Giveon}, {Becker}, {Helfand}, \&
  {White}}]{Giveon:2005}
{Giveon} U., {Becker} R.~H., {Helfand} D.~J., {White} R.~L., 2005, \aj, 130,
  156

\bibitem[{{Gold} {et~al}\mbox{.}(2011){Gold}, {Odegard}, {Weiland}, {Hill},
  {Kogut}, {Bennett}, {Hinshaw}, {Chen}, {Dunkley}, {Halpern}, {Jarosik},
  {Komatsu}, {Larson}, {Limon}, {Meyer}, {Nolta}, {Page}, {Smith}, {Spergel},
  {Tucker}, {Wollack}, \& {Wright}}]{Gold:2011}
{Gold} B. {et~al.}, 2011, \apjs, 192, 15

\bibitem[{{Gordon} \& {Cato}(1972)}]{Gordon:1972}
{Gordon} M.~A., {Cato} T., 1972, \apj, 176, 587

\bibitem[{{Green}(2009{\natexlab{a}})}]{Green:2009}
{Green} D.~A., 2009{\natexlab{a}}, 37, 45

\bibitem[{{Green}(2009{\natexlab{b}})}]{Green:2009b}
---, 2009{\natexlab{b}}, \mnras, 399, 177

\bibitem[{{Hart} \& {Pedlar}(1976)}]{Hart:1976}
{Hart} L., {Pedlar} A., 1976, \mnras, 176, 547

\bibitem[{{Haslam} \& {Osborne}(1987)}]{Haslam:1987}
{Haslam} C.~G.~T., {Osborne} J.~L., 1987, \nat, 327, 211

\bibitem[{{Helfand} {et~al}\mbox{.}(1992){Helfand}, {Zoonematkermani},
  {Becker}, \& {White}}]{Helfand:1992}
{Helfand} D.~J., {Zoonematkermani} S., {Becker} R.~H., {White} R.~L., 1992,
  \apjs, 80, 211

\bibitem[{{Jaffe} {et~al}\mbox{.}(2011){Jaffe}, {Banday}, {Leahy}, {Leach}, \&
  {Strong}}]{Jaffe:2011}
{Jaffe} T.~R., {Banday} A.~J., {Leahy} J.~P., {Leach} S., {Strong} A.~W., 2011,
  ArXiv e-prints

\bibitem[{{Kalberla} {et~al}\mbox{.}(2010){Kalberla}, {McClure-Griffiths},
  {Pisano}, {Calabretta}, {Alyson Ford}, {Lockman}, {Staveley-Smith}, {Kerp},
  {Winkel}, {Murphy}, \& {Newton-McGee}}]{Kalberla:2010}
{Kalberla} P.~M.~W. {et~al.}, 2010, ArXiv e-prints

\bibitem[{{Kaplan} {et~al}\mbox{.}(2002){Kaplan}, {Kulkarni}, {Frail}, \& {van
  Kerkwijk}}]{Kaplan:2002}
{Kaplan} D.~L., {Kulkarni} S.~R., {Frail} D.~A., {van Kerkwijk} M.~H., 2002,
  \apj, 566, 378

\bibitem[{{Lockman}(1976)}]{Lockman:1976}
{Lockman} F.~J., 1976, \apj, 209, 429

\bibitem[{{Mezger} \& {Henderson}(1967)}]{Mezger&Henderson:1967}
{Mezger} P.~G., {Henderson} A.~P., 1967, \apj, 147, 471

\bibitem[{{Mezger}(1978)}]{Mezger:1978}
{Mezger} P.~O., 1978, \aap, 70, 565

\bibitem[{{Miville-Desch{\^e}nes} \& {Lagache}(2005)}]{Miville-Deschenes:2005}
{Miville-Desch{\^e}nes} M., {Lagache} G., 2005, \apjs, 157, 302

\bibitem[{{Paladini} {et~al}\mbox{.}(2003){Paladini}, {Burigana}, {Davies},
  {Maino}, {Bersanelli}, {Cappellini}, {Platania}, \& {Smoot}}]{Paladini:2003}
{Paladini} R., {Burigana} C., {Davies} R.~D., {Maino} D., {Bersanelli} M.,
  {Cappellini} B., {Platania} P., {Smoot} G., 2003, \aap, 397, 213

\bibitem[{{Paladini}, {Davies} \& {DeZotti}(2004){Paladini}, {Davies}, \&
  {DeZotti}}]{Paladini:2004}
{Paladini} R., {Davies} R.~D., {DeZotti} G., 2004, \mnras, 347, 237

\bibitem[{{Planck Collaboration} {et~al}\mbox{.}(2011){Planck Collaboration},
  {Abergel}, {Ade}, {Aghanim}, {Arnaud}, {Ashdown}, {Aumont}, {Baccigalupi},
  {Balbi}, {Banday}, \& et~al.}]{Planck:2011a}
{Planck Collaboration} {et~al.}, 2011, ArXiv e-prints

\bibitem[{{Putman} {et~al}\mbox{.}(2002){Putman}, {de Heij}, {Staveley-Smith},
  {Braun}, {Freeman}, {Gibson}, {Burton}, {Barnes}, {Banks}, {Bhathal}, {de
  Blok}, {Boyce}, {Disney}, {Drinkwater}, {Ekers}, {Henning}, {Jerjen},
  {Kilborn}, {Knezek}, {Koribalski}, {Malin}, {Marquarding}, {Minchin},
  {Mould}, {Oosterloo}, {Price}, {Ryder}, {Sadler}, {Stewart}, {Stootman},
  {Webster}, {Wright}, \& et~al.}]{Putman:2002}
{Putman} M.~E. {et~al.}, 2002, \aj, 123, 873

\bibitem[{{Reich} \& {Reich}(1988)}]{Reich:1988}
{Reich} P., {Reich} W., 1988, \aaps, 74, 7

\bibitem[{{Reich} {et~al}\mbox{.}(1990){Reich}, {Fuerst}, {Reich}, \&
  {Reif}}]{Reich:1990a}
{Reich} W., {Fuerst} E., {Reich} P., {Reif} K., 1990, \aaps, 85, 633

\bibitem[{{Reich}, {Reich} \& {Fuerst}(1990){Reich}, {Reich}, \&
  {Fuerst}}]{Reich:1990}
{Reich} W., {Reich} P., {Fuerst} E., 1990, \aaps, 83, 539

\bibitem[{{Reynolds}(1991)}]{Reynolds:1991}
{Reynolds} R.~J., 1991, \apjl, 372, L17

\bibitem[{{Rohlfs} \& {Wilson}(2004)}]{Rohlfs:2004}
{Rohlfs} K., {Wilson} T.~L., 2004, {Tools of radio astronomy}, Rohlfs K.
  \&~Wilson T.~L., ed.

\bibitem[{{Shaver} {et~al}\mbox{.}(1983){Shaver}, {McGee}, {Newton}, {Danks},
  \& {Pottasch}}]{Shaver:1983}
{Shaver} P.~A., {McGee} R.~X., {Newton} L.~M., {Danks} A.~C., {Pottasch} S.~R.,
  1983, \mnras, 204, 53

\bibitem[{{Slee}(1995)}]{Slee:1995}
{Slee} O.~B., 1995, Australian Journal of Physics, 48, 143

\bibitem[{{Staveley-Smith} {et~al}\mbox{.}(1998){Staveley-Smith}, {Juraszek},
  {Koribalski}, {Ekers}, {Green}, {Haynes}, {Henning}, {Kesteven},
  {Kraan-Korteweg}, {Price}, \& {Sadler}}]{Staveley-Smith:1998}
{Staveley-Smith} L. {et~al.}, 1998, \aj, 116, 2717

\bibitem[{{Staveley-Smith} {et~al}\mbox{.}(2003){Staveley-Smith}, {Kim},
  {Calabretta}, {Haynes}, \& {Kesteven}}]{Staveley-Smith:2003}
{Staveley-Smith} L., {Kim} S., {Calabretta} M.~R., {Haynes} R.~F., {Kesteven}
  M.~J., 2003, \mnras, 339, 87

\bibitem[{{Staveley-Smith} {et~al}\mbox{.}(1996){Staveley-Smith}, {Wilson},
  {Bird}, {Disney}, {Ekers}, {Freeman}, {Haynes}, {Sinclair}, {Vaile},
  {Webster}, \& {Wright}}]{Staveley-Smith:1996}
{Staveley-Smith} L. {et~al.}, 1996, Publications of the Astronomical Society of
  Australia, 13, 243

\bibitem[{{Stil} {et~al}\mbox{.}(2006){Stil}, {Taylor}, {Dickey}, {Kavars},
  {Martin}, {Rothwell}, {Boothroyd}, {Lockman}, \&
  {McClure-Griffiths}}]{Stil:2006}
{Stil} J.~M. {et~al.}, 2006, \aj, 132, 1158

\end{thebibliography}

%%%%%%%%%%%%%%%%%%%%%%%%%%%%%%%%%%%%%%%%%%%%%%%%%%%%%%%%%%%%%%%%%%%%
%%%%%%%%%%%%%%%%%%%%%%%%%%%%%%%%%%%%%%%%%%%%%%%%%%%%%%%%%%%%%%%%%%%

\bsp % ``This paper has been produced using the ...''

\label{lastpage}

\end{document}